\documentclass[onecolumn, 12pt]{IEEEtran}
\usepackage{setspace}
\ifCLASSINFOpdf
   \usepackage[pdftex]{graphicx}
   \graphicspath{{../pdf/}{../jpeg/}}
   \DeclareGraphicsExtensions{.pdf,.jpeg,.png}
\else
   \usepackage[dvips]{graphicx}
   \graphicspath{{../eps/}}
   \DeclareGraphicsExtensions{.eps}
\fi
\usepackage{epstopdf}
\usepackage{amsthm}
\usepackage{amsmath}
\usepackage{bm}
\usepackage{stfloats}

\usepackage{subfigure}
\usepackage{caption}
\usepackage{algorithm}
\usepackage{algorithmic}
\usepackage{courier}
\usepackage{amssymb}
\usepackage{color}

\begin{document}
\captionsetup[figure]{labelfont={default},labelformat={default},labelsep=period,name={Fig.}}
\title{Cooperative Reflection and Synchronization Design for Distributed Multiple-RIS Communications}
%
%
%
\author{Yaqiong Zhao, \emph{Student Member, IEEE,}
       Wei Xu, \emph{Senior Member, IEEE,}\\
       Xiaohu You, \emph{Fellow, IEEE,}
       Ning Wang, \emph{Member, IEEE,}
       and Huan Sun, \emph{Member, IEEE}

 \thanks{Y. Zhao, W. Xu, and X. You are with the National Mobile Communications Research Lab, and Frontiers Science Center for Mobile Information Communication and Security, Southeast University, Nanjing 210096, China, and also with Purple Mountain Laboratories, Nanjing 211111, China (\{zhaoyaqiong, wxu, xhyu\}@seu.edu.cn).}
 \thanks{N. Wang is with the School of Information Engineering, Zhengzhou University, Zhengzhou 450001, China (ienwang@zzu.edu.cn).}
\thanks{H. Sun is with the Wireless Technology Laboratory, Huawei Technologies Co. Ltd, Shanghai 201206, China (sunhuan11@huawei.com).}
}

\maketitle

\begin{abstract}
To reap the promised gain achieved by distributed reconfigurable intelligent surface (RIS)-enhanced communications in a wireless network, timing synchronization among these metasurfaces is an essential prerequisite in practice. This paper proposes a unified framework for the joint estimation of the unknown timing offsets and the RIS channel parameters, as well as the design of cooperative reflection and synchronization algorithm for the distributed multiple-RIS communication. Considering that RIS is usually a passive device with limited capability of signal processing, the individual timing offset and channel gains of each hop of the RIS links cannot be directly estimated. To make the estimation tractable, we propose to estimate the cascaded channels and timing offsets jointly by deriving a maximum likelihood estimator. Furthermore, we theoretically characterize the Cram{\'e}r-Rao lower bound (CRLB) to evaluate the accuracy of this estimator. By using the proposed estimator and the derived CRLBs, an efficient resynchronization algorithm is devised jointly at the RISs and the destination to compensate the multiple timing offsets. Based on the majorization-minimization framework, the proposed algorithm admits semi-closed and closed form solutions for the RIS reflection matrices and the timing offset equalizer, respectively. Simulation results
verify that our theoretical analysis well matches the numerical tests and validate the effectiveness of the proposed resynchronization algorithm.

\end{abstract}


%
\IEEEpeerreviewmaketitle

\section{Introduction}

With the rapid development of emerging multimedia
applications such as virtual reality (VR) and augmented
reality (AR), there is an urgent demand to
support higher data rates in the fifth-generation (5G) cellular network and beyond [1]-[3]. To achieve this goal, there have emerged a number of physical
layer enhancements, including massive multiple-input multiple-output (MIMO), millimeter-wave (mmWave) communication, ultra-dense cloud radio access network (UD-CRAN) [3]-[5], etc. These techniques are anticipated to provide, all together, a 1000-fold increase
in the network capacity, by reaping joint benefits
of large-antenna gains, wide spectrum bands, and dense deployment of heterogeneous networks. However, these techniques often come with remarkable increase in hardware cost and power consumption, which raise serious concerns from the perspective of green and sustainable communications [6], [7].

Reconfigurable intelligent surface (RIS) is an emerging energy- and cost-efficient technology that is able to improve communication quality and extend coverage with significant reductions in the requirement of additional resources [8], [9]. Specifically, the RIS is a reprogrammable metasurface consisting of a large number of cost-effective passive reflecting elements. Each passive reflecting element is able to independently adjust the phase shift of the incoming signal under the control of a smart controller. By adjusting the phases
of the reflected signals according to the instantaneous wireless channel states, the received signal copies at the destination always add constructively. Thus, the received signal power is strengthened, that is, the RIS artificially creates favorable channel conditions and offers degrees-of-freedom (DoF) for performance improvement [5], [8], [9]. Compared with a typical amplify-and-forward (AF) relay, RIS consumes much less energy since it only reflects the signals passively like a mirror. It neither generates additional noise, nor imposes any necessity for power amplifiers or other active radio-frequency (RF) components. Moreover, RISs can be manufactured flexibly so that they can be easily integrated into existing communication environments, such as building facades and room ceilings [8]. Since RIS eliminates the use of RF sources and operates only in limited range, it can be densely deployed without considering complicated interference management among multiple passive RISs [10]. All these advantages have motivated extensive studies on RIS in both academia and industry.

Researches have investigated the deployment of RIS for both single-user communications [11]-[13] and multiuser scenarios [14]-[16]. In particular, the study in [11] investigated the problem of channel estimation for a RIS-assisted MIMO system and proposed a two-stage algorithm using sparse matrix factorization and matrix completion techniques. In [12], the transmit beamforming and RIS reflection were jointly optimized to maximize the received signal power, where the base station (BS) employed maximal-ratio transmission (MRT) and the RIS reflection pattern was optimized by using semidefinite relaxation (SDR). In [13], the achievable rate of a RIS-assisted downlink system was derived, which gives insights into the theoretical performance limit of RIS. Considering multiuser setups, a typical
on/off-based channel estimation framework was advocated in [14], where only one reflecting element is switched on in each training phase in order to obtain separate user channels. Besides, the problems of sum-rate and energy efficiency maximization were respectively considered in [15] and [16] for RIS-assisted multiuser systems.

To further unlock the potential of cooperative RISs, some recent works have focused on multiple-RIS-assisted systems [17]-[22]. In particular, deploying distributed RISs in a communication network helps to enhance the coverage and to achieve stable transmission. In [17] and [18], a typical weighted sum-rate maximization problem was studied for a multiple-RIS-assisted system, under the assumption of ideal and non-ideal hardware, respectively.  Furthermore, the authors in [19] and [20] focused on the statistical
characterization and performance analysis of systems aided by multiple RISs. The ergodic achievable rate for cooperative RISs was theoretically derived under Rician fading and Nakagami-$m$ fading, respectively. Beyond the above works, the use of RISs for improved energy efficiency has also been investigated in [21]. In [21], the authors attempted to maximize the energy efficiency of a wireless network deploying multiple RISs by optimizing the reflection coefficients of the RISs and dynamically controlling the on-off status of each RIS. Instead of considering the availability of perfect channel state information (CSI), the authors in [22] proposed a channel estimation protocol for the distributed RIS-assisted system using the Bayesian technique of minimum mean squared error (MMSE) estimation.

It is worth noting that existing works on multiple-RIS-assisted communications mostly assumed perfect timing synchronization among the RISs. However, it is challenging to achieve perfect timing synchronization in practice, if not impossible, for the multiple passive RISs. Owing to its nature as a passive device, RIS does not have any capability of implementing signal processing. Consequently, the individual timing offset of each hop of the links cannot be obtained directly by the RISs, and thus cannot be compensated by, e,g., analog delay components at each RIS circuit as in conventional relaying systems. More importantly, optimizing the system parameters based
on perfect timing synchronization would inevitably lead to performance degradation due to time dispersion and subsequent inter-symbol interference (ISI) [23], [24]. The previous work [25] optimized the cooperative RISs by taking into account this asynchronization, while perfect channel and timing information was assumed to be known as \emph{a prior}. However, this can be hardly realized even in a conventional communication system, not to mention in the system with passive RISs.

In this paper, we try to fill this gap by proposing a unified framework for the joint estimation
of timing offsets and channel status, as well as the design of cooperative reflection and synchronization algorithm at the RISs and the destination for the distributed multiple-RIS-assisted network. The main contributions of this paper are summarized as follows.
\begin{itemize}
  \item To the best of our knowledge, this is the first investigation on the joint estimation of timing offsets and RIS channel coefficients in a multiple-RIS-assisted communication system. Specifically, we propose a general RIS reflection pattern satisfying the unit-modulus constraints for the estimation rather than a simple on/off-based reflection pattern. By exploring the diagonal structure of the RIS reflection matrices, an efficient maximum likelihood estimator (MLE) is derived for the cascaded channels and timing offsets.
  \item For this multiple parameter estimation problem, closed-form Cram{\'e}r-Rao lower bounds (CRLBs) are derived and then utilized to evaluate the performance of the proposed estimator. As expected, simulation results demonstrate that the mean-squared error (MSE) of the estimate well approaches the CRLB. It verifies that the MLE asymptotically attains the optimum estimation performance as predicted by the CRLB [26].
  \item Building upon the parameter estimates and their corresponding CRLBs, we propose to minimize the MSE of the system over these estimation uncertainties by a joint optimization of RIS reflection matrices and a timing offset equalizer. This problem is nonconvex with a fourth-order polynomial objective function, block-diagonal constraints, and unit-modulus constraints. By reformulating the objective into a more tractable form, we devise a majorization-minimization (MM)-based algorithm to solve this problem. The RIS reflection matrices and the timing offset equalizer are fortunately obtained in semi-closed and closed forms, respectively. Convergence of the proposed algorithm is theoretically proved and an acceleration scheme is further presented.
\end{itemize}

The remainder of this paper is organized as follows. In Section \uppercase\expandafter{\romannumeral2}, the system model is introduced. In Section \uppercase\expandafter{\romannumeral3}, the problem of joint timing offset and cascaded RIS channel estimation is formulated and the corresponding theoretical analysis of the CRLB is also presented. In Section \uppercase\expandafter{\romannumeral4}, by using the proposed estimator and the derived CRLBs, we further optimize the RIS reflection matrices, jointly with the timing offset equalizer. Simulation results are presented in Section \uppercase\expandafter{\romannumeral5}, followed by concluding remarks in Section \uppercase\expandafter{\romannumeral6}.

\emph{Notations:} In this paper, ${\mathbb C}^{M\times N}$ and ${\mathbb R}^{M\times N}$ respectively denote the space of $M\times N$ complex and real matrices. Superscripts $(\cdot)^{T}$, $(\cdot)^{*}$, and $(\cdot)^{H}$ respectively denote the transpose, the conjugate, and the conjugate transpose operators. Operator diag$\left(  \cdot  \right)$ returns a diagonal matrix with the input as its elements and blk$\left[{\bf A},\cdots,{\bf B}\right]$ denotes the block-diagonal matrix with ${\bf A},\cdots,{\bf B}$ on its diagonal. Re$\{\cdot\}$ and Im$\{\cdot\}$ take the real part and the imaginary part of a complex quantity, respectively. vec$(\cdot)$ stands for the vectorization operation and $\frac{{\partial f}}{{\partial x}}$ is the partial derivative of $f$ with respect to $x$. Notation $\otimes$ means the Kronecker product. The operator tr$\{\cdot\}$ takes the trace of the input matrix and  $\mathbb{E}_x\{\cdot\}$ takes the expectation with respect to $x$. Notations $\|\cdot\|_1$, $\|\cdot\|_2$, and $\|\cdot\|_{F}$, respectively, are the $L_1$ norm, the $L_2$ norm, and the Frobenius norm, while $\lambda_{\text{max}}(\cdot)$ is the maximum eigenvalue of the input matrix. ${\cal {CN}}(0,\sigma^2)$ represents the complex Gaussian distribution with zero mean and variance $\sigma^2$. ${\bf I}_m$ denotes the $m\times m$ identity matrix and $\bf 0$ denotes an all-zero matrix. ${\cal O}(\cdot)$ and $o(\cdot)$ represent the standard big-O and little-o notations, respectively. Notation $\simeq $ means asymptotically equal to. The directional derivative of $f({\bf x})$ in the direction of vector $\bf d$ is given by $f'({\bf{x}};{\bf{d}})\triangleq \mathop {\lim }\limits_{\lambda  \to 0} \frac{{f({\bf{x}} + \lambda {\bf{d}}) - f({\bf{x}})}}{\lambda }$. Finally, arg$\{\cdot\}$ takes the argument of a complex value.

\section{System Model}
\begin{figure}[!t]
\centering 
\includegraphics[width=9cm,height=8.5cm]{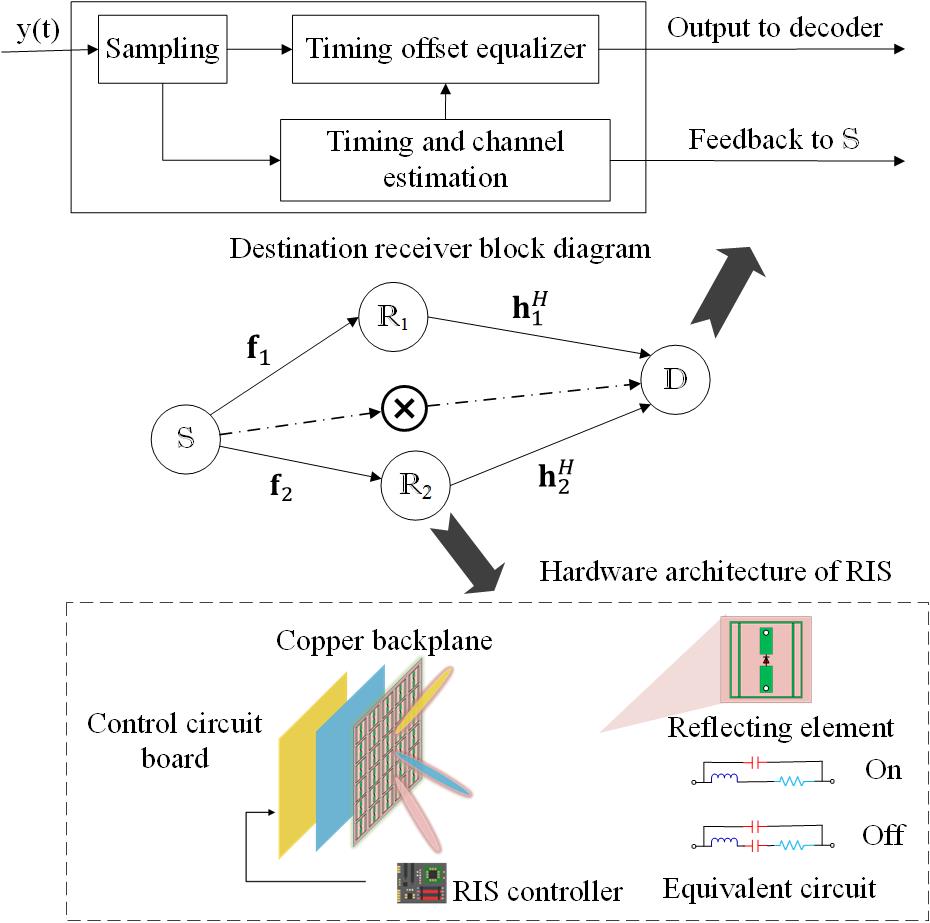} 
\caption*{\small{Fig. 1 System model for a multiple-RIS-assisted system.}}
\label{fig:0}       
\end{figure}
We consider a multiple-RIS-assisted downlink communication system as shown in Fig. 1. The system consists of a single-antenna source, ${\mathbb S}$, a single-antenna destination, ${\mathbb D}$, and $K$ distributed RISs, ${\mathbb R}_k$, $\forall k\in{\cal K}\triangleq\{1,2,\cdots,K\}$. Each RIS is equipped with $N$ passive reflecting elements. Reasonably, the signal reflected by the RIS more than once is neglected due to high penetration losses. Thanks to their passive nature, we assume that no direct communication links exist between the multiple RISs and the cooperation between them are realized under the control of ${\mathbb S}$  through RIS controllers, as adopted in the literature [18]-[22]. Moreover, direct links between ${\mathbb S}$ and ${\mathbb D}$ are negligible in the proposed model due to unfavorable propagation conditions as in [9], [16]. Generally, it happens more often that the received power of direct path is too weak to establish a communication link in future communication systems using high-frequency bands since the channels at these frequencies,  e.g., $30-100$ GHz, are significantly more vulnerable to severe path attenuation and signal blockages than that at sub-$6$ GHz frequencies [10], [27], [28].

Due to hardware inconsistencies and diverse RIS locations, the signals arriving at ${\mathbb D}$ from the RISs generally experience different propagation delays, and thus they usually arrive asynchronously at ${\mathbb D}$ [24]. In order to compensate the performance degradation caused by this asynchronization, a joint cooperative reflection and synchronization
design is motivated in this work. In practice, since the source ${\mathbb S}$  has no prior information of timing and channel states, a training phase before data transmission is usually helpful. In particular, the communication process includes the following two phases.
\begin{itemize}
  \item {\bf Training phase:} During the training phase, ${\mathbb S}$ sends pilot signals to ${\mathbb D}$ through the $K$ distributed passive RISs. After sampling, a joint estimation of the timing offsets and cascaded channels of all the RISs is performed at ${\mathbb D}$. The timing offset in our paper refers to the overall timing offset which consists of both the timing offset of the ${\mathbb S}$-${\mathbb R}_k$ link and that of the ${\mathbb R}_k$-${\mathbb D}$ link. After ${\mathbb D}$ feeds back the estimates of the timing offsets and the cascaded channels, these estimates are further utilized by ${\mathbb S}$ to determine the joint cooperative reflection and synchronization design.
  \item {\bf Data transmission phase:} After the training phase, ${\mathbb S}$ transmits the data sequence to the $K$ RISs. All the RISs cooperatively forward the incoming signals to ${\mathbb D}$ using the designed reflection matrices. The received signal at ${\mathbb D}$ is further processed by the proposed timing offset equalizer before the signals can be demodulated.
\end{itemize}

We assume a quasi-static flat-fading channel model for all the channels involved in the considered setup. Denote by ${\bf f}_k\in {\mathbb C}^{N\times 1}$ and ${\bf h}_k\in {\mathbb C}^{N\times 1}$ the channel from ${\mathbb S}$ to ${\mathbb R}_k$ and that from ${\mathbb D}$ to ${\mathbb R}_k, \forall k\in{\cal K}$. Define ${\bf W}_k= \text {diag}({\bm \theta _k}) \in {\mathbb C}^{N\times N}$ as the passive reflection matrix of ${\mathbb R}_k$, where ${\bm \theta _k}=\left[\theta_{k,1},\theta_{k,2},\cdots,\theta_{k,N}\right]^T$ with $\theta_{k,l}$ being the corresponding reflection coefficient introduced by the $l$th element of ${\mathbb R}_k$, $\forall l\in{\cal L}\triangleq\{1,2,\cdots,N\}$. Let ${\bf s}\triangleq[s(0),s(1),\cdots,s(L-1)]^T \in {\mathbb C}^{L\times 1}$ denote the symbol vector of modulated data transmitted by ${\mathbb S}$. Then, the continuous-time received signal at ${\mathbb D}$ for $0\leq t\leq L_oT$ is expressed as
\begin{equation}\label{eq:timedomin}
y(t)=\!\sum_{k=1}^{K}{\bf h}_k^H{\bf W}_k{\bf f}_k\!\!\!\!\sum_{i=-L_g}^{L_o+L_g-1}\!s(i) g\left(t-i T-\epsilon_{k} T\right)+v(t),
\end{equation}
where $g(t)$ is the transmitting pulse shaping filter, $\epsilon_k$ is the normalized timing offset of ${\mathbb R}_k$ over the symbol duration $T,~\forall k\in\cal{K}$, and $v(t)$ is the zero-mean circularly symmetric complex Gaussian noise with variance $\sigma^2$. Note that the length of the sequence ${\bf s}$ is taken to be $L=2L_g+L_o$, where $L_o$ and $L_g$ respectively stand for the length of observation interval and the effective duration of the tail of $g(t)$ on one side [24], [29].

Upon reception, the waveform $y(t)$ is oversampled at ${\mathbb D}$ by a factor $Q=T/T_s$, where $T_s$ is the sample interval. After stacking the $L_oQ$ samples of $y(t)$ into a vector, i.e., ${\bf y}\triangleq[y(0),y(T_s),\cdots,y(L_oQ-1)T_s]^T$, we have
\begin{equation}\label{eq:vector}
{\bf y}=\sum_{k=1}^{K} \left({\bf h}_k^{\text H}{\bf W}_k{\bf f}_{k}\right){\bf A}(\epsilon_k) {\bf s}+{\bf v},
\end{equation}
where
\begin{align}\label{eq:definitation}
\mathbf{A}(\epsilon_k)&\triangleq\left[\mathbf{a}_{-L_g}\left(\epsilon_{k}\right), \cdots,\mathbf{a}_{0}\left(\epsilon_{k}\right),\cdots, \mathbf{a}_{L_o+L_g-1}\left(\epsilon_{k}\right)\right]\nonumber\\
\mathbf{a}_{i}\left(\epsilon_{k}\right) &\triangleq \left[g\left(-i T-\epsilon_{k} T\right), g\left(-iT+T_s-\epsilon_{k} T\right),\cdots,g\left(-i T+\left(L_oQ-1\right) T_{s}-\epsilon_{k} T\right)\right]^{T}\nonumber\\
\mathbf v&\triangleq[v(0),v(T_s),\cdots,v(L_oQ-1)T_s]^T.\nonumber
\end{align}

In the next section, an estimation algorithm is first devised for the joint estimation of channel gains and timing offsets. Given the obtained estimates, a design of cooperative reflection and synchronization algorithm is then presented.
\section{Joint Timing and Channel Estimation}
Timing and channel estimation is of preliminary importance for the design of cooperative reflection at the RISs and the timing offset equalizer at the destination. However, it is difficult to explicitly acquire the separate channel coefficients and timing offsets
of each of the ${\mathbb S}$-${\mathbb R}_k$ link and the ${\mathbb R}_k$-${\mathbb D}$ link. That is because RIS is a passive antenna array which only reflects electromagnetic waves without any active RF components [10]. To overcome this difficulty, we propose to jointly estimate the cascaded channels and timing offsets of the multiple RISs in the training phase.
\subsection{Joint Timing and Channel Estimation Method}
We start with an equivalent reformulation of (\ref{eq:vector}) by using the diagonal structure of the RIS reflection matrices, ${\bf W}_k,~\forall k\in{\cal K}$. Define ${\bf s}_t$ as the pilot signals in the training phase. Let ${\bf M}(\bm\epsilon)\triangleq\left[{\bf A}(\epsilon_1){\bf s}_t,{\bf A}(\epsilon_2){\bf s}_t,\cdots,{\bf A}(\epsilon_K){\bf s}_t\right]\in{\mathbb C}^{L_oQ\times K}$, ${\bf H}_{\text{e}}\triangleq {\text{blk}}\left[{\bf h}_1^H{\text{diag}}({\bf f}_1),{\bf h}_2^H{\text{diag}}({\bf f}_2),\cdots,{\bf h}_K^H{\text{diag}}({\bf f}_K)\right]\in{\mathbb C}^{K\times NK}$, and ${\bm\theta}\triangleq\left[{\bm\theta}_1^T,{\bm\theta}_2^T,\cdots,{\bm\theta}_K^T\right]^T\in{\mathbb C}^{NK\times 1}$. The system model in (\ref{eq:vector}) during the training phase is concisely rewritten as
\begin{equation}\label{eq:estimation}
{\bf y} = {\bf M}(\bm\epsilon){\bf H}_{\text{e}}{\bm\theta}+{\bf v}.
\end{equation}
Assume that the training phase is divided into $M$ sub-phases. The channel and timing offsets remain unchanged throughout the training phase within a coherence time. From (\ref{eq:estimation}), the received training signal at ${\mathbb D}$ during the $m$th sub-phase, ${\bf y}[m]$, is expressed as
\begin{equation}\label{eq:ym}
{\bf y}[m] = {\bf M}(\bm\epsilon){\bf H}_{\text{e}}{\bm\theta}[m]+{\bf v}[m],
\end{equation}
where ${\bm\theta}[m]$ and ${\bf v}[m]$, respectively, correspond to the reflection coefficients of the RISs and the noise vector at the receiver at the $m$th subphase, $\forall m\in{\cal M}=\{1,2,\cdots,M\}$.

Now by stacking the received training signals in (\ref{eq:ym}) of all the $M$ sub-phases, the received signal equals
\begin{equation}\label{eq:yt1}
{\bf Y} = {\bf M}(\bm\epsilon){\bf H}_{\text{e}}{\bm\Phi}^T+{\bf V},
\end{equation}
where ${\bf Y}=\left[{\bf y}[1],{\bf y}[2],\cdots,{\bf y}[M]\right]$, ${\bm\Phi}= \left[{\bm\theta}[1],{\bm\theta}[2],\cdots,{\bm\theta}[M]\right]^T$, and ${\bf V}=\left[{\bf v}[1],{\bf v}[2],\cdots,{\bf v}[M]\right]$. By utilizing the equality of Kronecker product, i.e., $\text{vec}\left({\bf{AXB}}\right)=\left( {{{\bf{B}}^T} \otimes {\bf{A}}} \right)\text{vec}({\bf X})$, we further rewrite (\ref{eq:yt1}) as
\begin{equation}\label{eq:yt}
{\bf y}_t = \left({\bm\Phi}\otimes {\bf M}(\bm\epsilon)\right)\text{vec}({\bf H}_{\text{e}})+{\bf v}_t={\bf N}(\bm\epsilon){\bf h}_{\text{eq}}+{\bf v}_t,
\end{equation}
where ${\bf y}_t\triangleq\text{vec}({\bf Y})=\left[{\bf y}[1]^T,{\bf y}[2]^T,\cdots,{\bf y}[M]^T\right]^T$, ${\bf v}_t\triangleq\text{vec}({\bf V})=\left[{\bf v}[1]^T,{\bf v}[2]^T,\cdots,{\bf v}[M]^T\right]^T$, ${\bf h}_{\text{eq}}\triangleq\left[{\bf h}_1^H{\text{diag}}({\bf f}_1),{\bf h}_2^H{\text{diag}}({\bf f}_2),\cdots,{\bf h}_K^H{\text{diag}}({\bf f}_K)\right]^T$ represents the cascaded channels of all $K$ RISs, and
\begin{equation}\label{eq:Ne}
{\bf N}(\bm\epsilon)\triangleq\left[{\bm\Phi}_1\otimes\left({\bf A}(\epsilon_ 1){\bf s}_t\right),\cdots,{\bm\Phi}_K\otimes\left({\bf A}(\epsilon_K){\bf s}_t\right)\right],
\end{equation}
where ${\bm\Phi}=\left[{\bm\Phi}_1,{\bm\Phi}_2,\cdots,{\bm\Phi}_K\right]$. Note that in writing the second equality of (\ref{eq:yt}), we apply \emph {Lemma 1} in Appendix A to remove the zero elements of $\text{vec}({\bf H}_{\text{e}})$.

Note that the RIS reflection pattern, $\bm\Phi$, is a reflection pattern that satisfies the unit-modulus constraints, i.e., the reflecting elements are always switched on. Compared with the traditional on/off scheme in [14], this RIS reflection pattern achieves better estimation performance since the large beamforming gain of the RIS can be fully exploited with all the elements switched on. Without loss of generality, it is assumed that the matrix $\bm\Phi$ is chosen to satisfy ${\bm\Phi}{\bm\Phi}^H=NK{\bf I}_{M}$, which indicates that ${\bm\Phi}$ is a scaled unitary matrix [30].

On the other hand, in order to minimize the estimation errors, e.g., estimation MSE, the pilot signals should be properly designed jointly with the RIS reflection pattern [31], [32]. Since the focus of this paper is on the problem of timing resynchronization, ${\bf s}_t$ is chosen as white sequences so as to obtain good performance in terms of MSE and CRLB as verified via numerical tests [24]. Typically, this is a reasonable choice since white sequences cooperated with a scaled unitary matrix $\bm\Phi$ helps to increase the rank of ${\bf N}(\bm\epsilon)$ as observed from its definition in (\ref{eq:Ne}), thereby improving the estimation accuracy.

Having the equivalent model in (\ref{eq:yt}), we are ready to estimate the cascaded channels, ${\bf h}_{\text{eq}}$, and the timing offsets, ${\bm\epsilon}\triangleq[\epsilon_1,\epsilon_2,\cdots,\epsilon_K]^T$, based on the design philosophy of MLE. Assume that the noise vectors during the training phase are uncorrelated and follow the same distribution, i.e., ${\bf v}_t\sim{\cal {CN}}({\bf 0},\sigma_t^2{\bf I}_{ML_oQ})$. Then the joint likelihood function of the timing offsets and the cascaded channels is expressed as
\begin{equation}\label{eq:ML}
P\left({\bf y}_t|{\bm\epsilon},{\bf h}_{\text{eq}}\right)={\left( {\pi \sigma _t^2} \right)^{ - ML_oQ}}\exp \left\{ - \frac{\left\| {\bf y}_t - {\bf N}(\bm\epsilon){\bf h}_{\text{eq}} \right\|^2}{\sigma _t^2} \right\}.
\end{equation}
Given that system parameters are constants, it is straightforward to show that maximizing the above likelihood function is equivalent to minimizing the following cost function
\begin{equation}\label{eq:costfunction}
{\Lambda }\left({\bm\epsilon},{\bf h}_{\text {eq}} \right)\triangleq\left\| {\bf y}_t - {\bf N}(\bm\epsilon){\bf h}_{\text{eq}} \right\|^2.
\end{equation}
Then the problem of joint estimation of timing offsets and channel coefficients is formulated as
\begin{equation}\label{eq:estimator}
(\hat{\bm\epsilon}, {\hat {\bf h}_{\text{eq}}})=\text{arg}\mathop {\text{min}}\limits_{{\bm\epsilon},{\bf h}_{\text{eq}}} {\Lambda}\left( {\bm\epsilon},{\bf h}_{\text{eq}} \right).
\end{equation}
For any ${\bm\epsilon}$, it can be shown that the optimal estimate of ${\bf h}_{\text{eq}}$ follows
\begin{equation}\label{eq:heq}
{\hat {\bf h}}_{\text{eq}} = {\left( {\bf N}(\bm\epsilon)^H{\bf N}(\bm\epsilon) \right)^{ - 1}}{\bf N}(\bm\epsilon)^H{\bf y}_t.
\end{equation}
By substituting (\ref{eq:heq}) back into (\ref{eq:costfunction}), the log-likelihood cost function is obtained as
\begin{equation}\label{eq:deltak}
{\Lambda}\left( {{\bm\epsilon}} \right)\triangleq {\left\|{\bf P}({\bm\epsilon}){\bf y}_t \right\|^2},
\end{equation}
where ${\bf P}({\bm\epsilon})\triangleq{\bf I}-{\bf N}(\bm\epsilon)\left({\bf N}(\bm\epsilon)^H{\bf N}(\bm\epsilon)\right)^{-1}{\bf N}(\bm\epsilon)^H$. From (\ref{eq:deltak}), the optimal estimate of $\bm\epsilon$ is expressed as
\begin{equation}\label{eq:epsilon}
\hat{\bm\epsilon} =\text{arg}\mathop {\text{min}}\limits_{{\bm\epsilon}} {\Lambda }\left({\bm\epsilon} \right),
\end{equation}
and the optimal estimate of ${\hat {\bf h}}_{\text{eq}}$ is obtained by plugging $\hat{\bm\epsilon}$ into (\ref{eq:heq}). To solve problem (\ref{eq:epsilon}), it requires an exhaustive search over a multidimensional
space, imposing very high computational complexity. To overcome this challenge, we exploit the technique of alternating projection to reduce the $K$-dimensional minimization in (\ref{eq:epsilon}) to a series of low complexity $1$-D minimization problems [33]. By applying alternating projection, the proposed MLE is summarized in Algorithm 1, where ${\Lambda }\left({\bm\epsilon}_{(t,k)},{\epsilon}_k \right)$ is used to indicate the functional dependence of ${\Lambda }\left( {{\bm\epsilon}} \right)$ on ${\left[{\epsilon}_{1}^{(t+1)},\cdots,{\epsilon}_{k-1}^{(t+1)},{\epsilon}_k,{\epsilon}_{k+1}^{(t)}\cdots,{\epsilon}_{K}^{(t)}\right]^T} $.
\begin{algorithm}[t]
\caption{Proposed MLE using Alternating Projection}
\label{alg:Framwork}
\begin{algorithmic}[1] 
\STATE Set $t=0$ and initialize ${\bm\epsilon}.$
\REPEAT
\FOR{$k=1$ to $K$}
\STATE${\bm\epsilon}_{(t,k)} ={\left[{\epsilon}_{1}^{(t+1)},\cdots,{\epsilon}_{k-1}^{(t+1)},{\epsilon}_{k+1}^{(t)}\cdots,{\epsilon}_{K}^{(t)}\right]^T} $;
\STATE${\epsilon}_{k}^{(t+1)}=\text{arg}\mathop {\text{min}}\limits_{{\epsilon}_k} {\Lambda }\left({\bm\epsilon}_{(t,k)},{\epsilon}_k \right)$;
\ENDFOR
\STATE$t \rightarrow  t+1$;
\UNTIL{convergence}
\STATE$\hat{\bm\epsilon} ={\left[{\epsilon}_{1}^{(t)},{\epsilon}_{2}^{(t)},\cdots,{\epsilon}_{K}^{(t)}\right]} $;
\STATE${\hat {\bf h}}_{\text{eq}} = {\left( {\bf N}(\hat{\bm\epsilon})^H{\bf N}(\hat{\bm\epsilon}) \right)^{ - 1}}{\bf N}(\hat{\bm\epsilon})^H{\bf y}_t.$
\end{algorithmic}
\end{algorithm}
\subsection{CRLB Analysis}
The accuracy of the proposed joint estimation of timing offsets and channel parameters depends largely on the noise. That is to express
\begin{align}\label{eq:errors}
{\bm\epsilon}&= \hat{\bm\epsilon}+{\bm\delta}\left(\bm\epsilon\right),\nonumber\\
{\bf h}_{\text{eq}}&= {\hat{\bf h}}_{\text{eq}}+{\bm\delta}\left({\bf h}_{\text{eq}}\right),
\end{align}
where ${\bm\delta}\left(\bm\epsilon\right)\in{\mathbb R}^{K\times 1}$ and ${\bm\delta}\left({\bf h}_{\text{eq}}\right)\in{\mathbb C}^{NK\times 1}$ are the random estimation errors of the corresponding estimates of $\bm\epsilon$ and ${\bf h}_{\text{eq}}$, respectively. Typically, the CRLB serves as a benchmark for evaluating unbiased estimators. It defines a lower bound on the statistical variance of any unbiased estimator. In this subsection, we present a detailed analysis on the CRLBs of the timing and channel parameter estimates to verify the effectiveness of the proposed estimator.

Recall the signal model in (\ref{eq:yt}) during the training phase. Let ${\bm\xi}\triangleq \left[{\bm\epsilon}^T,\text{Re}\{{\bf h}_{\text{eq}}\}^T,\text{Im}\{{\bf h}_{\text{eq}}\}^T\right]^T$ and denote ${\bm\mu}\triangleq {\bf N}({\bm\epsilon}){\bf h}_{\text{eq}}$. According to [26], [34], the Fisher Information Matrix (FIM), denoted by $\bf J$, of ${\bm\xi}$ is defined as
\begin{equation}\label{eq:fisher0}
{\bf J}=\frac{2}{\sigma _t^2}{\text{Re}}\left\{\frac{{\partial {\bm\mu }^H}}{{\partial {\bm\xi}}}\frac{{\partial {\bm\mu }}}{{\partial {\bm\xi}^T}}\right\}.
\end{equation}
In order to have each block of $\bf J$, the following readily obtained relations are necessary:
\begin{align}\label{eq:deritives}
\frac{{\partial {\bm\mu }^H}}{{\partial {\bm\epsilon}}}&= {\bf H}_{\text e}^*{\bf N}_{\text d}^H(\bm\epsilon),\nonumber\\
\frac{{\partial {\bm\mu }^H}}{{\partial \text{Re}\{{\bf h}_{\text{eq}}\}}}&={\bf N}^H(\bm\epsilon),\nonumber\\
\frac{{\partial {\bm\mu }^H}}{{\partial \text{Im}\{{\bf h}_{\text{eq}}\}}}&=-\jmath{\bf N}^H(\bm\epsilon),
\end{align}where ${\bf N}_{\text d}(\bm\epsilon)\triangleq\left[{\bm\Phi}_1\otimes\left({\bf D}(\epsilon_ 1){\bf s}_t\right),\cdots,{\bm\Phi}_K\otimes\left({\bf D}(\epsilon_K){\bf s}_t\right)\right]$ with ${\bf D}(\epsilon_i)\triangleq\frac{{\partial {\bf A}(\epsilon_i)}}{{\partial {\epsilon _i}}}$. By substituting (\ref{eq:deritives}) into (\ref{eq:fisher0}) and after some straightforward manipulations, the expression for the FIM is obtained as
\begin{equation}\label{eq:fisher}
\begin{array}{l}
{\bf J} = \frac{2}{{\sigma _t^2}}\times\left[ {\begin{array}{*{20}{c}}
{{\text{Re}}\left\{{\bf H}_{\text e}^*{\bf N}_{\text d}(\bm\epsilon)^H{\bf N}_{\text d}(\bm\epsilon){\bf H}_{\text e}^T\right\} }&{{\text{Re}}\left\{{\bf H}_{\text e}^*{\bf N}_{\text d}(\bm\epsilon)^H{\bf N}(\bm\epsilon)\right\} }&{{-\text{Im}}\left\{{\bf H}_{\text e}^*{\bf N}_{\text d}(\bm\epsilon)^H{\bf N}(\bm\epsilon)\right\} }\\
{{\text{Re}}\left\{{\bf N}(\bm\epsilon)^H{\bf N}_{\text d}(\bm\epsilon){\bf H}_{\text e}^T\right\}}&{{\text{Re}}\left\{{\bf N}(\bm\epsilon)^H{\bf N}(\bm\epsilon)\right\}}&{{-\text{Im}}\left\{{\bf N}(\bm\epsilon)^H{\bf N}(\bm\epsilon)\right\}}\\
{{\text{Im}}\left\{{\bf N}(\bm\epsilon)^H{\bf N}_{\text d}(\bm\epsilon){\bf H}_{\text e}^T\right\}}&{{\text{Im}}\left\{{\bf N}(\bm\epsilon)^H{\bf N}(\bm\epsilon)\right\}}&{{\text{Re}} \left\{{\bf N}(\bm\epsilon)^H{\bf N}(\bm\epsilon)\right\}}
\end{array}} \right]
\end{array}.
\end{equation}
By inverting $\bf J$, the CRLBs for the timing and channel parameter estimates are presented in the following theorem.

\emph{Theorem 1:}
The CRLB matrices for the proposed estimator are established as follows
\begin{align}\label{eq:theorem1}
{\bf C}(\bm\epsilon)&=\frac{\sigma _t^2}{2}\left({\text{Re}}\left\{{\bf H}_{\text e}^*{\bf N}_{\text d}(\bm\epsilon)^H{\bf P}(\bm\epsilon){\bf N}_{\text d}(\bm\epsilon){\bf H}_{\text e}^T\right\}\right)^{-1},\nonumber\\
{\bf C}\left({\bf h}_{\text{eq}}\right)&=\!\frac{\sigma _t^2}{2}\!\left(\!2\left({\bf N}(\bm\epsilon)^H{\bf N}(\bm\epsilon)\right)^{-1}\!\!\!+\!\left({\bf N}(\bm\epsilon)^H{\bf N}(\bm\epsilon)\right)^{-1}\!{\bf N}(\bm\epsilon)^H{\bf N}_{\text d}(\bm\epsilon){\bf H}_{\text e}^T\left({\text{Re}}\left\{{\bf H}_{\text e}^*{\bf N}_{\text d}(\bm\epsilon)^H{\bf P}(\bm\epsilon){\bf N}_{\text d}(\bm\epsilon){\bf H}_{\text e}^T\right\}\right)^{-1}\right.\nonumber\\
&\quad\left.\times{\bf H}_{\text e}^*{\bf N}_{\text d}(\bm\epsilon)^H{\bf N}(\bm\epsilon)\left({\bf N}(\bm\epsilon)^H{\bf N}(\bm\epsilon)\right)^{-1}\right).
\end{align}
\begin{proof}
See Appendix B.
\end{proof}
With the derived CRLBs in \emph{Theorem 1}, we then apply the asymptotic properties of the MLE to the parameter estimates. From [26, \emph{Theorem 7.1}], it is theoretically proven that MLE is asymptotically unbiased and it asymptotically attains the CRLB. Accordingly, for the proposed estimator, the estimation errors of the timing and channel parameters, i.e., ${\bm\delta}\left(\bm\epsilon\right)$ and ${\bm\delta}\left({\bf h}_{\text{eq}}\right)$, follow an asymptotical multivariate Gaussian distribution as
\begin{align}\label{eq:distribution}
{\bm\delta}\left(\bm\epsilon\right)&\sim{\cal {N}}({\bf 0},{\bf C}(\bm\epsilon)),\nonumber\\
{\bm\delta}\left({\bf h}_{\text{eq}}\right)&\sim{\cal {CN}}\left({\bf 0},{\bf C}\left({\bf h}_{\text{eq}}\right)\right).
\end{align}
Equipped with the parameter estimates and their associated distribution, it is now possible to design a cooperative reflection and synchronization algorithm to compensate the multiple timing offsets at the destination.
\section{Cooperative Reflection and Synchronization
Design}
In the previous section, we obtain the estimates of the timing offsets and cascaded channels. The
remaining task is to utilize these estimates, as well as their corresponding statistical estimation error information, to compensate the asynchronization. In the following, we propose a practical timing resynchronization design for the RIS reflection matrices, ${\bm \Psi}$, and the timing offset equalizer, $\bf G$, at ${\mathbb D}$, by considering both asynchronization and estimation errors.

Recall the general model in (\ref{eq:vector}). The received signal ${\bf y}_d$ during the data transmission phase can be rewritten in a matrix form as
\begin{equation}\label{eq:matrix}
{\bf y}_d={\bf A}(\bm\epsilon){\bf H}{\bf W}{\bf F}{\bf s}_d+{\bf v},
\end{equation}
where $\mathbf{H}\triangleq\text{blk}\left[\mathbf h_{1}^{H}, \mathbf h_{2}^{H},\cdots, \mathbf h_{K}^{H}\right] \otimes{\mathbf I_L}$, $\mathbf{F} \triangleq\left[\mathbf f_{1}^{T}, \mathbf f_{2}^{T},\cdots, \mathbf f_{K}^{T}\right]^{T} \otimes{\mathbf I_L}$, $\mathbf{W} \triangleq \text{blk}\left[\mathbf W_{1}, \mathbf W_{2},\cdots, \mathbf W_{K}\right] \otimes{\mathbf I_L}$, and ${\bf s}_d\in{\mathbb C}^{L\times 1}$ is the transmitted signal at ${\mathbb S}$ during the transmission phase with normalized power, i.e., $\mathbb{E}\left\{{\bf s}_d{\bf s}_d^H\right\}={\bf I}_L$. Considering $\bf W$ is diagonal, the signal model in (\ref{eq:matrix}) is further rearranged as
\begin{equation}\label{eq:yd}
{\bf y}_d={\bf A}(\bm\epsilon){\bm \Psi}{\bf H}_{\text{eq}}{\bf s}_d+{\bf v},
\end{equation}
where ${\bm \Psi}\triangleq\text{blk}\left[{\bm\theta}_1^T,{\bm\theta}_2^T,\cdots,{\bm\theta}_K^T\right]\otimes{\bf I}_L$ and ${\bf H}_{\text{eq}}\triangleq {\bf h}_{\text{eq}}\otimes{\bf I}_L$.
\subsection{Problem Reformulation}
To detect ${\bf y}_d$ in (\ref{eq:yd}), we propose to jointly optimize the designs of $\bm \Psi$ and $\bf G$ by minimizing the MSE under unit-modulus constraints. It yields
\begin{subequations}\label{eq:problem}
\begin{align}
({\cal P}_1)\quad&{\mathop{\text{min}}\limits_{_{{\bm \Psi} ,{\bf G}}}}\quad\text{MSE}({\bm\Psi},{\bf G})\triangleq\mathbb{E}\left\{ {\left\|{\bf Gy}_d - {\bf T}({\bm{\eta }}){\bf s}_d \right\|^2 }\right\}\label{eq:problema}\\
&~\text{s.t.}\quad{\bm \Psi}=\text{blk}\left[{\bm\theta}_1^T,{\bm\theta}_2^T,\cdots,{\bm\theta}_K^T\right]\otimes{\bf I}_L,\label{eq:problemb}\\
&~\quad\quad\left|\theta_{k,l}\right|=1,\quad{\forall k\in \cal K},~{\forall l\in\cal L}\label{eq:problemc}
\end{align}
\end{subequations}
where ${\bf T}(\bm{\eta })$ is an $L_0\times L$
circulant matrix which is interpreted as a windowing operation selecting the length-$L_0$ block of the data for detection. The first row of ${\bf T}(\bm{\eta })$ is ${\bm\eta}^T$ and
\begin{equation}\label{eq:eta}
{\bm\eta}\triangleq\left[{\bf R}_{\text {g}}\left(-L_g\right), \cdots, {\bf R}_{\text {g}}(0), \cdots, {\bf R}_{\text {g}}\left(L_{g}\right) ~ \mathbf{0}_{1 \times\left(L_0-1\right)}\right]^T,
\end{equation}
where ${\bf R}_{\text {g}}(\tau )$ is the autocorrelation function of $g(t)$ sampled at $t=\tau T$.  Vector $\bm\eta$ denotes the ideally sampled waveform without ISI after matched filtering.

After some basic mathematical manipulations on the objective function of $({\cal P}_1)$ in Appendix C and by exploiting (\ref{eq:distribution}), we have the objective function in (\ref{eq:problema}) expressed as
\begin{align}\label{eq:MSE11}
\text{MSE}({\bm \Psi},{\bf G})&={\text{tr}}\left\{{\bf G}{\bf A}(\hat{\bm\epsilon}){\bm \Psi}{\bf R}_{{\bm\delta}({\hat{\bf H}_{\text{eq}}})}{\bm\Psi}^H{\bf A}(\hat{\bm\epsilon})^H{\bf G}^H\right\}+{\text{tr}}\left\{{\bf G}{\bf R}_v{\bf G}^H\right\}\nonumber\\
&~-2{\text{Re}}\!\left\{{\text{tr}}\!\left\{{\bf G}{\bf A}(\hat{\bm\epsilon}){\bm \Psi}\hat{\bf H}_{\text{eq}}{\bf T}^H({\bm{\eta }})\right\}\right\}+{\text{tr}}\left\{{\bf T}({\bm{\eta }}){\bf T}^H({\bm{\eta }})\right\}.
\end{align}
where $\hat{\bf H}_{\text{eq}}=\hat{\bf h}_{\text{eq}}\otimes{\bf I}_L$, ${\bf R}_{{\bm\delta}({\hat{\bf H}_{\text{eq}}})}={\hat{\bf H}_{\text{eq}}}{\hat{\bf H}^H_{\text{eq}}}+{\bf C}(\hat{\bf h}_{\text{eq}})\otimes{\bf I}_L$, and ${\bf R}_v\triangleq\mathbb{E}_{\bf v}\left\{{\bf v}{\bf v}^H\right\}$.

While (\ref{eq:problema}) has been reformulated into the closed form expression in (\ref{eq:MSE11}) by considering the estimation errors, problem $({\cal P}_1)$ is still challenging to be solved due to the coupling effects between the optimization variables and the block-diagonal and unit-modulus constraints imposed on the RIS reflection matrices. In the sequel, we present an efficient method to tackle these challenges and obtain high-quality suboptimal solutions to the optimization problem in $({\cal P}_1)$.

To begin with, we introduce the following lemma to convert problem $({\cal P}_1)$ to an equivalent problem for the sake of tractability.

\emph{Lemma 3:} The original problem in $({\cal P}_1)$ is equivalent to the following minimization problem
\begin{subequations}\label{eq:problem1}
\begin{align}
({\cal P}_2)\quad\quad\quad&{\mathop{\text{min}}\limits_{_{{\bm \Theta} ,{\bf G}}}}\quad\text{MSE}({\bm\Theta},{\bf G})\label{eq:problem1a}\\
&~\text{s.t.}\quad{\bm \Theta}={\bm\theta}^T\otimes{\bf I}_{L_oQ},\label{eq:problem1b}\\
&\quad\quad~\left|\theta_{k,l}\right|=1,\quad{\forall k\in \cal K},~{\forall l\in\cal L}\label{eq:problem1c}
\end{align}
\end{subequations}
wherein $\text{MSE}({\bm\Theta},{\bf G})$ is given by
\begin{equation}\label{eq:MSE33}
\text{MSE}({\bm \Theta},{\bf G})\!=\!{\text{tr}}\left\{{\bf G}{\bm\Theta}{\bm\Xi}{\bm\Xi}^H{\bm\Theta}^H{\bf G}^H\right\}+{\text{tr}}\left\{{\bf G}{\bf R}_v{\bf G}^H\right\}-2{\text{Re}}\left\{{\text{tr}}\left\{{\bf G}{\bm\Theta}{\bm\Pi}{\bf T}^H\!({\bm{\eta }})\right\}\right\}+{\text{tr}}\left\{{\bf T}({\bm{\eta }}){\bf T}^H({\bm{\eta }})\right\},\nonumber
\end{equation}
and
\begin{align*}
\!\begin{split}
\left\{ \begin{array}{l}
\!\!{\bm\Xi}\triangleq\left[{\bf R}_1^T\otimes{\bf A}(\hat\epsilon_1)^T\!,{\bf R}_2^T\otimes{\bf A}(\hat\epsilon_2)^T\!,\cdots,{\bf R}_K^T\otimes{\bf A}(\hat\epsilon_K)^T\right]^T\!\!\\
\!\!{\bm\Pi}\triangleq\left[\hat{\bf h}_{\text{eq},1}^T\!\otimes\!{\bf A}(\hat\epsilon_1)^T\!\!,\hat{\bf h}_{\text{eq},2}^T\!\otimes\!{\bf A}(\hat\epsilon_2)^T\!\!,\cdots\!,\hat{\bf h}_{\text{eq},K}^T\!\otimes\!{\bf A}(\hat\epsilon_K)^T\right]^T\!\!\!\!,\!\!\!
\end{array} \right.
\end{split}
\end{align*}
with ${\bf R}_{{\bm\delta}({\hat{\bf h}_{\text{eq}}})}={\hat{\bf h}_{\text{eq}}}{\hat{\bf h}^H_{\text{eq}}}+{\bf C}(\hat{\bf h}_{\text{eq}})$ and ${\bf R}_{{\bm\delta}({\hat{\bf h}_{\text{eq}}})}^{1/2}=\left[{\bf R}_1^T,{\bf R}_2^T,\cdots,{\bf R}_K^T\right]^T$, where $\hat{\bf h}_{\text{eq},k}\in{\mathbb C}^{N\times 1}$ represents the estimate of the cascaded channel corresponding to ${\mathbb R}_k$.
\begin{proof}
See Appendix D.
\end{proof}
Next, we deal with the coupling effect between the two optimization variables, $\bf G$ and $\bm\Theta$. Since it is easily verified that $\text{MSE}({\bm\Theta},{\bf G})$ is convex with respect to $\bf G$, we differentiate $\text{MSE}({\bm\Theta},{\bf G})$ with respect to $\bf G$ and set the derivative to zero. It admits a unique solution as a function of $\bm\Theta$ as follows
\begin{equation}\label{eq:equalizer}
{\bf G}={\bf T}(\bm{\eta }){\bm\Pi}^H{\bm\Theta}^H\left({\bm\Theta }{\bm\Xi}{\bm\Xi}^H{\bm\Theta}^H+{\bf R}_v \right)^{ - 1}.
\end{equation}
Substituting (\ref{eq:equalizer}) into the objective function of (\ref{eq:problem1}), we obtain the MSE expression with respect to only $\bm\Theta$, which is given by
\begin{align}\label{eq:MSE4}
\text{MSE}({\bm \Theta})&={\text{tr}}\left\{{\bf T}(\bm{\eta }){\bm\Pi}^H{\bm\Theta}^H\left({\bm\Theta }{\bm\Xi}{\bm\Xi}^H{\bm\Theta}^H+{\bf R}_v \right)^{ - 1}{\bm\Theta}{\bm\Xi}{\bm\Xi}^H{\bm\Theta}^H\left({\bm\Theta }{\bm\Xi}{\bm\Xi}^H{\bm\Theta}^H+{\bf R}_v \right)^{ - 1}{\bm\Theta}{\bm\Pi}{\bf T}^H(\bm{\eta })\right\}\nonumber\\
&\quad+{\text{tr}}\left\{{\bf T}(\bm{\eta }){\bm\Pi}^H{\bm\Theta}^H\left({\bm\Theta }{\bm\Xi}{\bm\Xi}^H{\bm\Theta}^H+{\bf R}_v \right)^{ - 1}{\bf R}_v\left({\bm\Theta }{\bm\Xi}{\bm\Xi}^H{\bm\Theta}^H+{\bf R}_v \right)^{ - 1}{\bm\Theta}{\bm\Pi}{\bf T}^H(\bm{\eta })\right\}\nonumber\\
&\quad-2{\text{tr}}\left\{{\bf T}(\bm{\eta }){\bm\Pi}^H{\bm\Theta}^H\left({\bm\Theta }{\bm\Xi}{\bm\Xi}^H{\bm\Theta}^H+{\bf R}_v \right)^{ - 1}{\bm\Theta}{\bm\Pi}{\bf T}^H({\bm{\eta }})\right\}+{\text{tr}}\left\{{\bf T}({\bm{\eta }}){\bf T}^H({\bm{\eta }})\right\}\nonumber\\
&=-{\text{tr}}\left\{{\bf T}(\bm{\eta }){\bm\Pi}^H{\bm\Theta}^H\left({\bm\Theta }{\bm\Xi}{\bm\Xi}^H{\bm\Theta}^H+{\bf R}_v \right)^{ - 1}{\bm\Theta}{\bm\Pi}{\bf T}^H({\bm{\eta }})\right\}+{\text{tr}}\left\{{\bf T}({\bm{\eta }}){\bf T}^H({\bm{\eta }})\right\}.
\end{align}
By safely dropping the constant term in (\ref{eq:MSE4}), the minimization problem $({\cal P}_2)$ in (\ref{eq:problem1}) is equivalently reformulated as the following maximization problem:
\begin{align}\label{eq:problem2}
({\cal P}_3)\quad\quad\quad&{\mathop{\text{max}}\limits_{_{{\bm \Theta} }}}\quad\overline{\text{MSE}}({\bm\Theta})\nonumber\\
&~\text{s.t.}\quad\text{(\ref{eq:problem1b})},\text{(\ref{eq:problem1c})},
\end{align}
where
\begin{equation}\label{eq:MSE_}
\overline{\text{MSE}}({\bm\Theta})\!\triangleq\!{\text{tr}}\!\left\{\!{\bf T}(\bm{\eta }){\bm\Pi}^H{\bm\Theta}^H\!\!\left({\bm\Theta }{\bm\Xi}{\bm\Xi}^H{\bm\Theta}^H\!\!\!+\!\!{\bf R}_v \right)^{ - 1}\!\!\!{\bm\Theta}{\bm\Pi}{\bf T}^H({\bm{\eta }})\!\right\}\nonumber.
\end{equation}

Up to now, we have reformulated the original problem $({\cal P}_1)$ into the equivalent problem in $({\cal P}_3)$. Unfortunately, the equivalent problem in (\ref{eq:problem2}) is still difficult to solve due to the following reasons. Firstly, the objective function $\overline{\text{MSE}}({\bm\Theta})$ is a fourth-order polynomial function of $\bm\Theta$. Secondly, the optimization variable $\bm\Theta$ is constrained to be sparse, i.e., ${\bm\Theta}={\bm\theta}^T\otimes{\bf I}_{L_oQ}$. Thirdly, the non-zero elements in $\bm\Theta$ are restricted to be unit-modulus with the hardware-constrained reflection elements of the RISs. In order to tackle this nonconvex optimization problem in $({\cal P}_3)$,
we devise an efficient algorithm based on the MM framework in the next subsection.
\subsection{Design of RIS Reflection Matrices}
The main idea of the MM framework is to minimize a sequence of surrogate functions with the same constraints as the original problem iteratively to well approximate the original problem [35], [36]. For instance, consider a minimization problem
\begin{equation}\label{eq:problemmm}
{\mathop{\text{min}}\limits_{_{{\bf x} }}}\quad f({\bf x}),\quad\text{s.t.}\quad {\bf x}\in{\cal X},
\end{equation}
and the feasible ponits $\{{\bf x}_{(t)}\}\in{\cal X}$ for (\ref{eq:problemmm}). The MM technique minimizes $f({\bf x})$ by minimizing a sequence of surrogate functions, $f\left({\bf x}|{\bf x}_{(t)}\right), t=0,1,\cdots$, satisfying the following three conditions [36]:
\begin{enumerate}
  \item $f\left({\bf x}|{\bf x}_{(t)}\right)\geq f({\bf x})$ for every feasible $\bf x$,
  \item $f\left({\bf x}_{(t)}|{\bf x}_{(t)}\right)= f({\bf x}_{(t)})$,
  \item ${\bf x}_{(t+1)}\in \text{arg}{\mathop{\text{min}}\limits_{_{{\bf x}\in{\cal X} }}}\quad f\left({\bf x}|{\bf x}_{(t)}\right)$.
\end{enumerate}
Then it can be readily proved that
\begin{equation}\label{eq:convergence}
f({\bf x}_{(t+1)})\leq f\left({\bf x}_{(t+1)}|{\bf x}_{(t)}\right)\leq f\left({\bf x}_{(t)}|{\bf x}_{(t)}\right)=f({\bf x}_{(t)}).
\end{equation}
As a result, the sequence of the solutions obtained in each iteration ultimately leads to a monotonically nonincreasing objective function value $\left\{f({\bf x}_{(t)}),~t=1,2,\cdots\right\}$. On the other hand, it is known that the objective value is bounded below. Then it is concluded that the sequence of objective values generated by the MM method always converge to a stationary point of the original problem.

For the problem $({\cal P}_3)$ in (\ref{eq:problem2}), the MM technique is employed as described in the following. To begin with, by denoting ${\bf X}= {\bm\Theta }{\bm\Xi}{\bm\Xi}^H{\bm\Theta}^H+{\bf R}_v$ and exploiting \emph{Lemma 4} in Appendix A, the objective of $({\cal P}_3)$, $\overline{\text{MSE}}({\bm\Theta})={\text{tr}}\left\{{\bf T}(\bm{\eta }){\bm\Pi}^H{\bm\Theta}^H{\bf X}^{ - 1}{\bm\Theta}{\bm\Pi}{\bf T}^H({\bm{\eta }})\right\}$, is jointly convex with respect to $\{{\bm\Theta},{\bf X}\}$. According to [37], a convex function is lower-bounded by its supporting hyperplane. It allows us to construct a surrogate function of $\overline{\text{MSE}}({\bm\Theta})$ as
\begin{align}\label{eq:MSEg1}
\overline{\text{MSE}}(\bm\Theta)&\geq\overline{\text{MSE}}({\bm\Theta}_{(t)})+2\text{Re}\left\{\text{tr}\left\{{\bm\Pi }{\bf T}^H({\bm\eta}){\bf F}_{(t)}^H{\bm\Theta}\right\}\right\}-\text{tr}\left\{{\bf F}_{(t)}^H{\bm\Theta}{\bm\Xi}{\bm \Xi}^H{\bm\Theta}^H{\bf F}_{(t)}\right\}+{\text {constant}}\nonumber\\
&\triangleq g_{\text{MSE}}\left({\bm\Theta},{\bm\Theta}_{(t)}\right),
\end{align}
where ${\bf F}_{(t)}\triangleq{\bf X}_{(t)}^{-1}{\bm\Theta}_{(t)}{\bm\Pi}{\bf T}^H({\bm\eta})$ with ${\bf X}_{(t)}={\bm\Theta }_{(t)}{\bm\Xi}{\bm\Xi}^H{\bm\Theta}_{(t)}^H+{\bf R}_v$ and the equality is achieved at ${\bm\Theta}={\bm\Theta}_{(t)}$. Hence, to solve the problem $({\cal P}_3)$ in (\ref{eq:problem2}), it can be sufficient to iteratively solve the following problem
\begin{align}\label{eq:problem3}
({\cal P}_4)\quad\quad\quad&{\mathop{\text{max}}\limits_{_{{\bm \Theta} }}}\quad g_{\text{MSE}}\left({\bm\Theta},{\bm\Theta}_{(t)}\right)\nonumber\\
&~\text{s.t.}\quad\text{(\ref{eq:problem1b})},\text{(\ref{eq:problem1c})},
\end{align}
While problem $({\cal P}_4)$ in (\ref{eq:problem3}) provides a surrogate
function with respect to $\bm\Theta$, it does not lead to a tractable formulation. Therefore, we introduce a further step by applying \emph{Lemma 5} in Appendix A to provide guidance in admitting a closed-form solution for $\bm\Theta$.

Applying \emph{Lemma 5} to $\text{tr}\left\{{\bf F}_{(t)}^H{\bm\Theta}{\bm\Xi}{\bm \Xi}^H{\bm\Theta}^H{\bf F}_{(t)}\right\}$, it is not hard to get
\begin{equation}\label{eq:MSEg2}
-\text{tr}\left\{{\bf F}_{(t)}^H{\bm\Theta}{\bm\Xi}{\bm \Xi}^H{\bm\Theta}^H{\bf F}_{(t)}\right\}\!\geq-\lambda_{(t)}\|{\bm\Theta}\|^2+\text{constant}+\!2\text{Re}\left\{\text{tr}\!\left\{\left(\lambda_{(t)}{\bm\Theta}_{(t)}- {\bf F}_{(t)}{\bf F}_{(t)}^H{\bm\Theta}_{(t)}{\bm\Xi}{\bm\Xi}^H\right)^H{\bm\Theta}\right\}\right\},
\end{equation}
where $\lambda_{(t)}\triangleq\|{\bm\Xi}{\bm\Xi}^H\|_1\|{\bf F}_{(t)}{\bf F}_{(t)}^H\|_1$. By plugging (\ref{eq:MSEg2}) into (\ref{eq:MSEg1}), it
leads to a further minorization to $\overline{\text{MSE}}({\bm\Theta})$ as
\begin{align}\label{eq:MSEg3}
\overline{\text{MSE}}(\bm\Theta) &\geq 2\text{Re}\left\{\text{tr}\left\{\left(\lambda_{(t)}{\bm\Theta}_{(t)}-{\bf F}_{(t)}{\bf F}_{(t)}^H{\bm\Theta}_{(t)}{\bm\Xi}{\bm\Xi}^H+{\bf F}_{(t)}{\bf T}(\bm\eta){\bm\Pi}^H\right)^H{\bm\Theta}\right\}\right\}+{\text {constant}}\nonumber\\
&\triangleq g\left({\bm\Theta}, {\bm\Theta}_{(t)}\right),
\end{align}
where we exploit the fact that $\|{\bm\Theta}\|_F^2=NKL_oQ$ (note that ${\bm\Theta}={\bm\theta}^T\otimes{\bf I}_{L_oQ}$). Thus far, $g\left({\bm\Theta}, {\bm\Theta}_{(t)}\right)$ can be regarded as a suitable surrogate
function for $\overline{\text{MSE}}(\bm\Theta)$ to employ the MM method. In other words, each iteration of the MM method requires to solve the following problem
\begin{align}\label{eq:problem4}
({\cal P}_5)\quad\quad\quad&{\mathop{\text{max}}\limits_{_{{\bm\Theta}}}}\quad g\left({\bm\Theta}, {\bm\Theta}_{(t)}\right) \nonumber\\
&~\text{s.t.}\quad\text{(\ref{eq:problem1b})},\text{(\ref{eq:problem1c})},
\end{align}
from which the optimal $\bm\Theta$ is obtained in the following theorem.

\emph{Theorem 2:} For any feasible ${\bm\Theta}_{(t)}$, the optimal solution for problem $({\cal P}_5)$ in (\ref{eq:problem4}) is
\begin{equation}\label{eq:solution}
{\theta}_{k,l}=e^{\jmath{\text{arg}}\left({\text{tr}}\left({\bf B}_{(t)}[k,l]\right)\right)},
\end{equation}
in which ${\bf B}_{(t)}[k,l]$ is a submatrix of ${\bf B}_{(t)}\triangleq\lambda_{(t)}{\bm\Theta}_{(t)}-{\bf F}_{(t)}{\bf F}_{(t)}^H{\bm\Theta}_{(t)}{\bm\Xi}{\bm\Xi}^H+{\bf F}_{(t)}{\bf T}(\bm\eta){\bm\Pi}^H$ containing the $\left[(kN-N+l-1)L_oQ+1\right]$th to $\left[(kN-N+l)L_oQ\right]$th columns for all $k\in{\cal K}$ and $l\in{\cal L}$.
\begin{proof} Note that ${\bm{\Theta }}={\bm\theta}^T\otimes {\bf I}_{L_oQ}$ and $\bm\theta={\left[ {\bm \theta _1^T, \cdots ,\bm \theta _K^T} \right]^T}$. By removing the constant term in (\ref{eq:MSEg3}), problem $({\cal P}_5)$ in (\ref{eq:problem4}) can be equivalently recast as
\begin{align}\label{eq:problem5}
({\cal P}_6)\quad\quad&{\mathop{\text{max}}\limits_{_{{\theta}_{k,l}}}}~ \sum_{k,l}^{}2\text{Re}\left\{\text{tr}\left\{{\bf B}^H_{(t)}[k,l]\right\}{\theta}_{k,l}\right\}\nonumber\\
&~\text{s.t.}\quad\text{(\ref{eq:problem1c})}.
\end{align}
Considering that $\{\theta_{k,l}\}$'s are independent variables, the objective in (\ref{eq:problem5}) is maximized when the phases of $\{\theta_{k,l}\}$'s are aligned with those of $\left\{{\text{tr}}\left({\bf B}_{(t)}[k,l]\right)\right\}$'s, which completes the proof.
\end{proof}

We summarize the procedure of the proposed cooperative reflection and synchronization
design in Algorithm 2.
\begin{algorithm}[t]
\caption{Cooperative Reflection
and Synchronization Design for Problem (\ref{eq:problem})}
\label{alg:Framwork}
\begin{algorithmic}[1] 
\STATE Set $t=0$ and initialize ${\theta}_{k,l}^{(t)},~\forall k\in{\cal K},~\forall l\in{\cal L}.$
\REPEAT
\STATE${\bm{\Theta }}_{(t)} ={\left[{\theta}_{1,1}^{(t)},\cdots,{\theta}_{1,N}^{(t)},\cdots,{\theta}_{K,N}^{(t)}\right]} \otimes {\bf I}_{L_oQ}$;
\STATE${\bf F}_{(t)}=\left({\bm\Theta }_{(t)}{\bm\Xi}{\bm\Xi}^H{\bm\Theta}_{(t)}^H+{\bf R}_v\right)^{-1}{\bm\Theta}_{(t)}{\bm\Pi}{\bf T}^H({\bm\eta})$;
\STATE$\lambda_{(t)}\triangleq\|{\bm\Xi}{\bm\Xi}^H\|_1\|{\bf F}_{(t)}{\bf F}_{(t)}^H\|_1$;
\STATE${\bf B}_{(t)}=\lambda_{(t)}{\bm\Theta}_{(t)}-{\bf F}_{(t)}{\bf F}_{(t)}^H{\bm\Theta}_{(t)}{\bm\Xi}{\bm\Xi}^H+{\bf F}_{(t)}{\bf T}(\bm\eta){\bm\Pi}^H$;
\STATE${\theta}_{k,l}^{(t+1)}=e^{\jmath{\text{arg}}\left({\text{tr}}\left\{{\bf B}[k,l]\right\}\right)},\forall k\in{\cal K},~\forall l\in{\cal L};$
\STATE$t \rightarrow  t+1$;
\UNTIL{convergence}
\STATE${\bm\Theta}={\bm\Theta}_{(t)}, {\bf G}={\bf T}(\bm{\eta }){\bm\Pi}^H{\bm\Theta}^H\left({\bm\Theta }{\bm\Xi}{\bm\Xi}^H{\bm\Theta}^H+{\bf R}_v \right)^{ - 1}.$
\end{algorithmic}
\end{algorithm}
\subsection{Algorithm Convergence and Acceleration}
Considering that Algorithm 2 is essentially based on the MM framework, its convergence can be rigorously proved. Briefly, the objective value of the original problem (\ref{eq:problem}) is non-increasing after each iteration of Algorithm 2 and eventually converges. In particular, we present the following theorem to validate the convergence for Algorithm 2.

\emph{Theorem 3:} By iteratively solving the sequence of problems $({\cal P}_5)$ in Algorithm 2, we get a series of points $\left\{{\bm\Theta}_{(t)},~t=0,1,\cdots\right\}$,
wherein each convergence point is a local optimum of the original problem $({\cal P}_1)$.
\begin{proof}
See Appendix E.
\end{proof}
Regarding the computational complexity, Algorithm 2 requires the computation of a matrix inversion and several matrix multiplications at each iteration. Hence, by considering popular Gaussian eliminations, the computational complexity per iteration of Algorithm 2 is in the order of ${\cal O}\left((NKL_oQ)^3\right)$.

Besides, it is worth noting that the objective function in (\ref{eq:problem2}) was minorized twice in deriving Algorithm 2, i.e., (\ref{eq:MSEg1}) and (\ref{eq:MSEg3}). This sometimes leads to a loose surrogate function and the performance of Algorithm 2 is susceptible to the slow convergence. For this reason, we employ an accelerated scheme based on the squared iterative method (SQUAREM) to improve both the convergence and the efficiency of Algorithm 2, without compromising on stability and feasibility [39]. Owing to its continuation property, SQUAREM enables a simple steplength backtracking strategy to diminish all the error components and thus can obtain global convergence. Note that SQUAREM is particularly appealing in solving high-dimensional problems when compared with other accelerators such as
quasi-Newton method with their simplicity and minimal storage requirements, which is exactly needed in massive MIMO and RIS-assisted systems.

By defining MMupdate$\left(\cdot\right)$ as performing Steps 3-7 of Algorithm 2 collectively, the accelerated method is presented in Algorithm 3. Specifically, the steplength $\alpha$ is calculated based on the Cauchy-Barzilai-Borwein method to improve the convergence rate in case the problem gets ill-conditioned [39]. A back-tracking strategy is employed in Step 8 to Step 10 in order to guarantee the monotonicity of the iterates generated by Algorithm 3. As for more details of the convergence and optimality analysis of the SQUAREM, please refer to [39].
\begin{algorithm}[t]
\caption{Accelerated Scheme for Algorithm 2.}
\label{alg:Framwork2}
\begin{algorithmic}[1] 
\STATE Set $t=0$ and initialize ${\theta}_{k,l}^{(t)},~\forall k\in{\cal K},~\forall l\in{\cal L}.$
\REPEAT
\STATE$\overline{\bm\theta}_1 ={\text{MMupdate}}\left({\bm\theta}_{(t)}\right),~\overline{\bm\theta}_2 ={\text{MMupdate}}\left(\overline{\bm\theta}_1\right)$;
\STATE${\bf r}=\overline{\bm\theta}_1-{\bm\theta}_{(t)},~{\bf v}=\overline{\bm\theta}_2-\overline{\bm\theta}_1-{\bf r}$;
\STATE Compute the step length $\alpha=-\frac{\|{\bf r}\|}{\|{\bf v}\|}$;
\STATE${\bm\theta}=e^{{\jmath}\text{arg}\left({\bm\theta_{(t)}}-2\alpha{\bf r}+\alpha^2{\bf v}\right)}$;
\STATE${\bm{\Theta }} ={\bm\theta}^T \otimes {\bf I}_{L_oQ}$;
\WHILE{$\text{MSE}\left(\bm\Theta\right)>\text{MSE}\left(\bm\Theta_{(t)}\right)$}
\STATE$\alpha \leftarrow (\alpha-1)/2$ and go to step 6;
\ENDWHILE
\STATE${\bm\theta_{(t+1)}}=\bm\theta$;
\STATE$t \rightarrow  t+1$;
\UNTIL{convergence}
\STATE${\bm\Theta}={\bm\Theta}_{(t)}, {\bf G}={\bf T}(\bm{\eta }){\bm\Pi}^H{\bm\Theta}^H\left({\bm\Theta }{\bm\Xi}{\bm\Xi}^H{\bm\Theta}^H+{\bf R}_v \right)^{ - 1}.$
\end{algorithmic}
\end{algorithm}
\section{Simulation Results}
In the following, we present numerical simulation results to validate our analysis as well as the benefits of the proposed schemes. In all the simulations, we consider a training
sequence ${\bf s}_t$ with $L_o = 12$ and $L_g=4$ and a typical square
root raised-cosine pulse shaping filter $g(t)$ with a roll-off factor 0.22 and normalized energy $\int_{ - \infty }^\infty  {{g^2}(t)} \text{d}t = 1$ [40]. The oversampling ratio $Q$ is set as $Q=2$. The signal-to-noise ratio (SNR) is defined as $\text{SNR}=\frac{E_s}{{{\sigma ^2}}}$ where $E_s$ denotes the average transmitted signal power. During the training phase, $M$ is chosen as $NK$ in order to achieve a complete estimation of all the $NK$ channel coefficients. The timing offsets are assumed to be uniformly distributed over the range $( -1, 1)$. Without loss of generality, we fix $E_s=1$. For the RIS, we consider a uniform rectangular array (URA) with $N= N_xN_y$, where $N_x$ and $N_y$ respectively stand for the number of elements in the horizontal axis and the vertical axis. Unless specified otherwise, we adopt $N_x=4$ in the simulations.
\subsection{Rayleigh Fading Channels}
In this subsection, we evaluate the performance of our proposed algorithm for estimation and timing resynchronization over Rayleigh fading channels.
\begin{figure}
\begin{minipage}[t]{0.5\linewidth}
\centering
\includegraphics[width=3.5in]{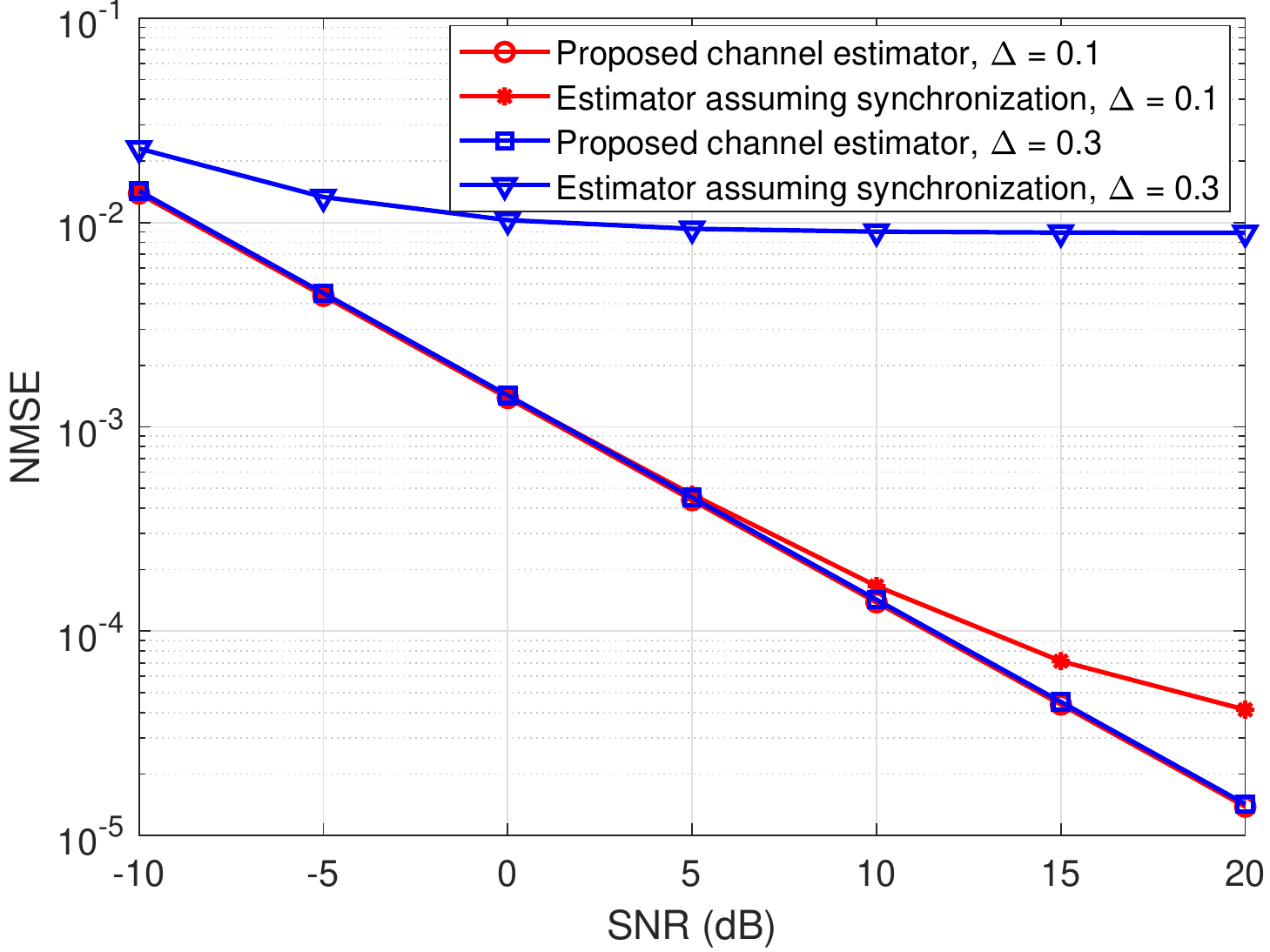}
\caption*{\small{Fig. 2  The effects of asynchronization on channel estimation, $\Delta_k\leq 0.1$ and $\Delta_k\leq 0.3$ for $K=2$ and $N=16$.}}
\label{fig:2}
\end{minipage}%
\begin{minipage}[t]{0.5\linewidth}
\centering
\includegraphics[width=3.5in]{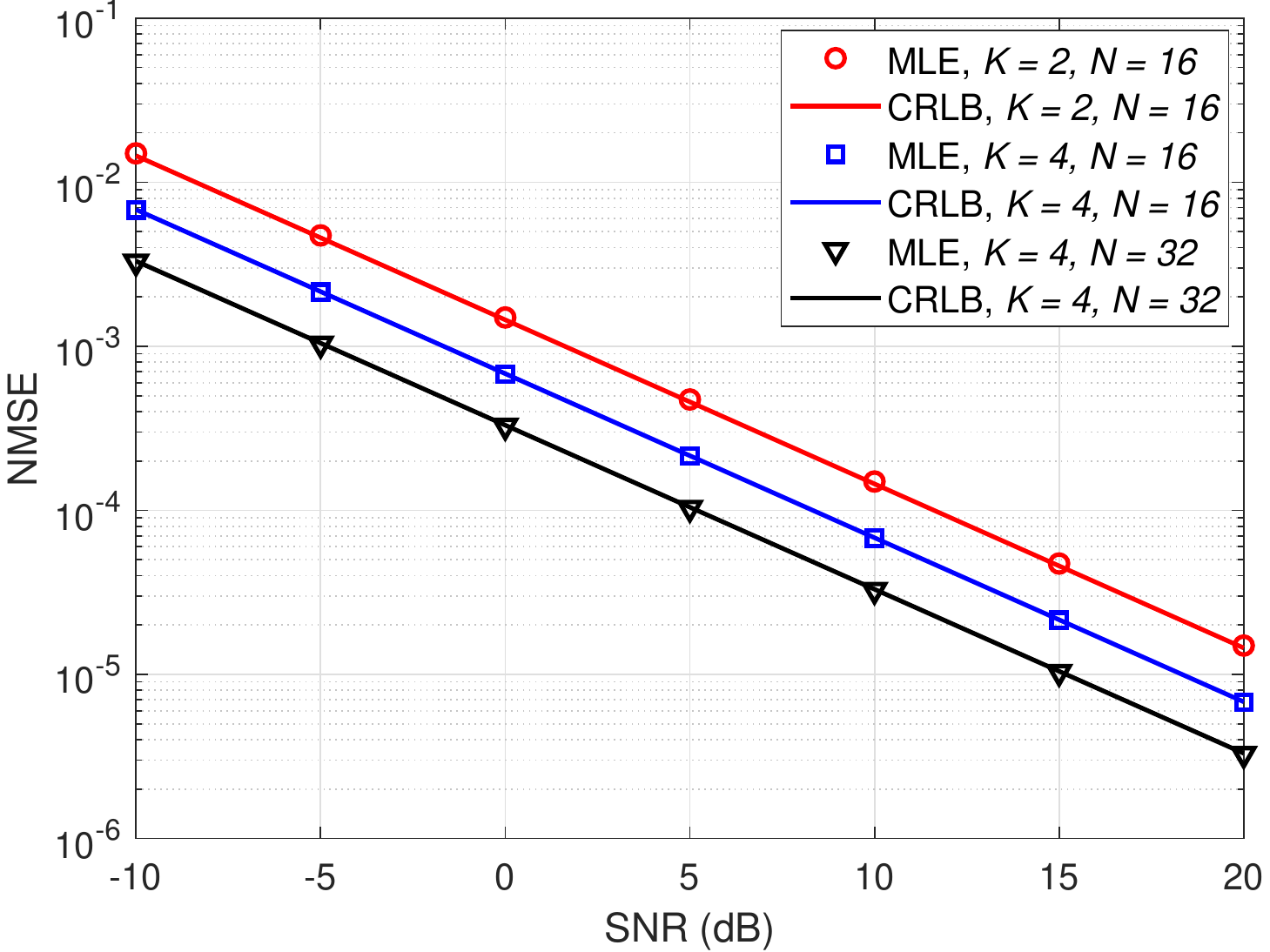}
\caption*{\small{Fig. 3  Estimation performance of the proposed MLE for the cascaded channels ${\bf h}_{\text{eq}}$ over Rayleigh channels.}}
\label{fig:3}
\end{minipage}
\end{figure}
In order to demonstrate the effects of asynchronization on channel estimation, we first examine in Fig. 2 the estimation errors of the cascaded channels under different levels of timing misalignment. The timing offsets are represented as $\epsilon_k=\epsilon_0+\Delta_k, \forall k\in\cal K$, where $\epsilon_k$ denotes the common offset and $\epsilon_k$ includes the travel delay difference corresponding to different RISs. Obviously, when the timing asynchronization is relatively mild, i.e., $\Delta_k\leq0.1$, the performance of the benchmark scheme which naively assumes perfect timing synchronization among the RISs overlaps with the performance of our proposed estimator at low SNRs while it begins to decline at high SNRs. This is due to the fact that the estimation performance is more affected by the noises at low SNR regions. On the contrary, when the system is under severe timing misalignment $\Delta_k\leq0.3$, the proposed estimator outperforms the benchmark scheme by orders of magnitude at all SNRs since the asynchronization contribution has severely dominated the system performance. Therefore, it is of great importance to take this timing asynchronization into the channel estimation of multiple-RIS systems.

In Fig. 3, the performance of our proposed estimation method and the derived CRLB in terms of the channel parameters are presented under different system setups. The normalized MSE (NMSE) of the cascaded channels and the associated CRLB are plotted as a function of SNR, where markers correspond to the estimation NMSE and solid lines represent the corresponding CRLB. It is observed that, for all SNRs, the performance of the proposed channel estimator matches well with the theoretical CRLB, validating the superiority of our proposed estimator and the accuracy of the derived CRLB. It also verifies that the proposed MLE can asymptotically attain the optimum estimation performance as predicted by the CRLB [26].

In Fig. 4, the timing estimation NMSE and the corresponding CRLB are plotted. It can be seen that for all cases, the NMSEs for timing estimates and their corresponding CRLBs coincide in high SNR region, indicating that the MLE solution provides very close accuracy to the theoretical CRLB. However, it is also observed that the timing estimation NMSEs fall below their associated CRLBs at low SNR regions. Strictly speaking, the CRLBs are not valid
bounds at low SNR regions and under fading environment for timing parameters since the derivation of CRLB does not utilize any prior information on the timing offset, i.e., the given finite range on the timing offsets. As a consequence, the CRLB that simply assumes the parameter as a real number with infinite range becomes inapplicable in this situation and this phenomenon can also be observed in [24]. As for channel parameters, the estimation errors can be well characterized by the CRLB since there is no range limit for any channel realization, as shown in Fig. 3.
\begin{figure}
\begin{minipage}[t]{0.5\linewidth}
\centering
\includegraphics[width=3.5in]{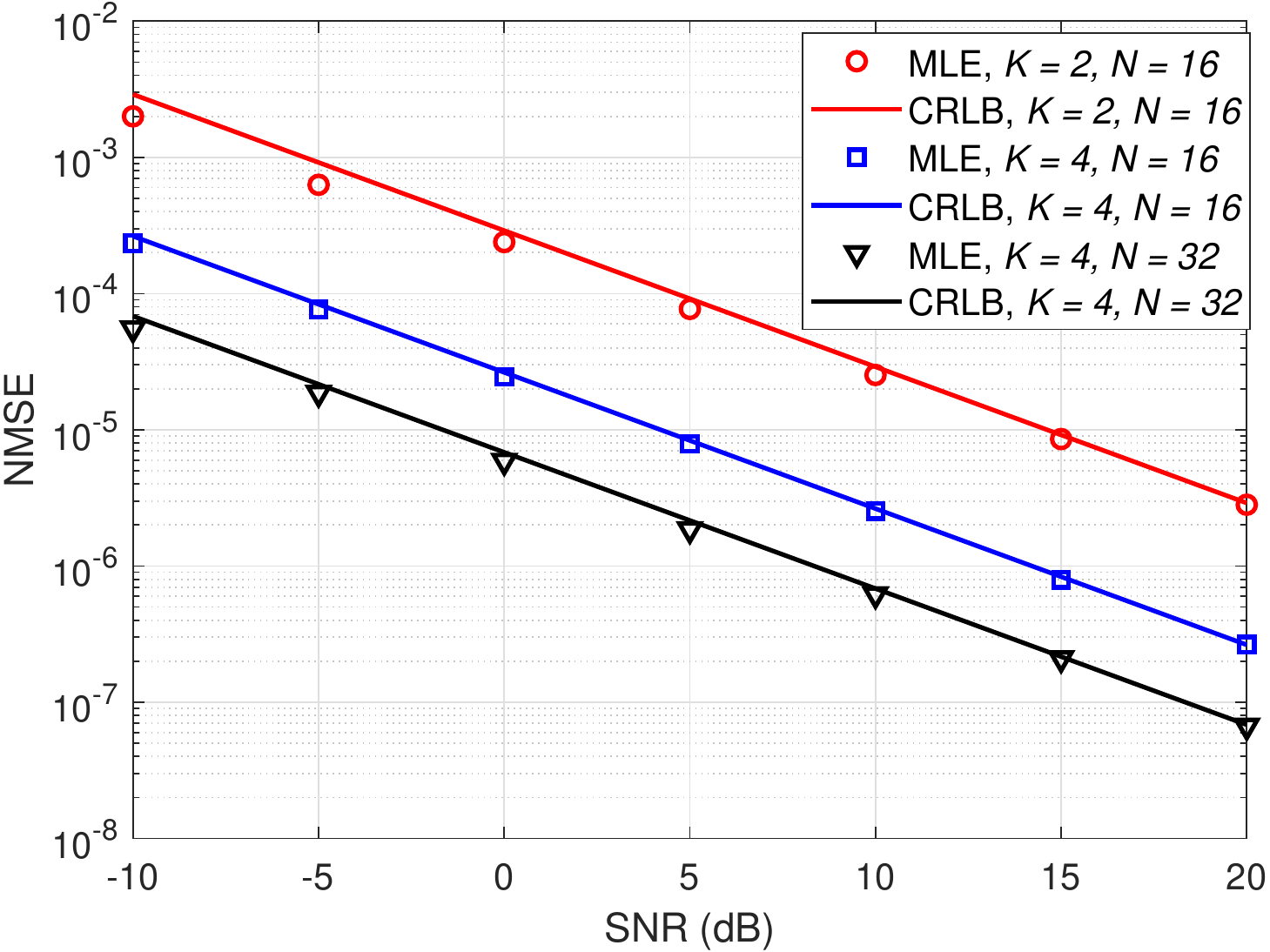}
\caption*{\small{Fig. 4  Estimation performance of the proposed MLE for the timing offsets $\bm\epsilon$ over Rayleigh channels.}}
\label{fig:2}
\end{minipage}%
\begin{minipage}[t]{0.5\linewidth}
\centering
\includegraphics[width=3.5in]{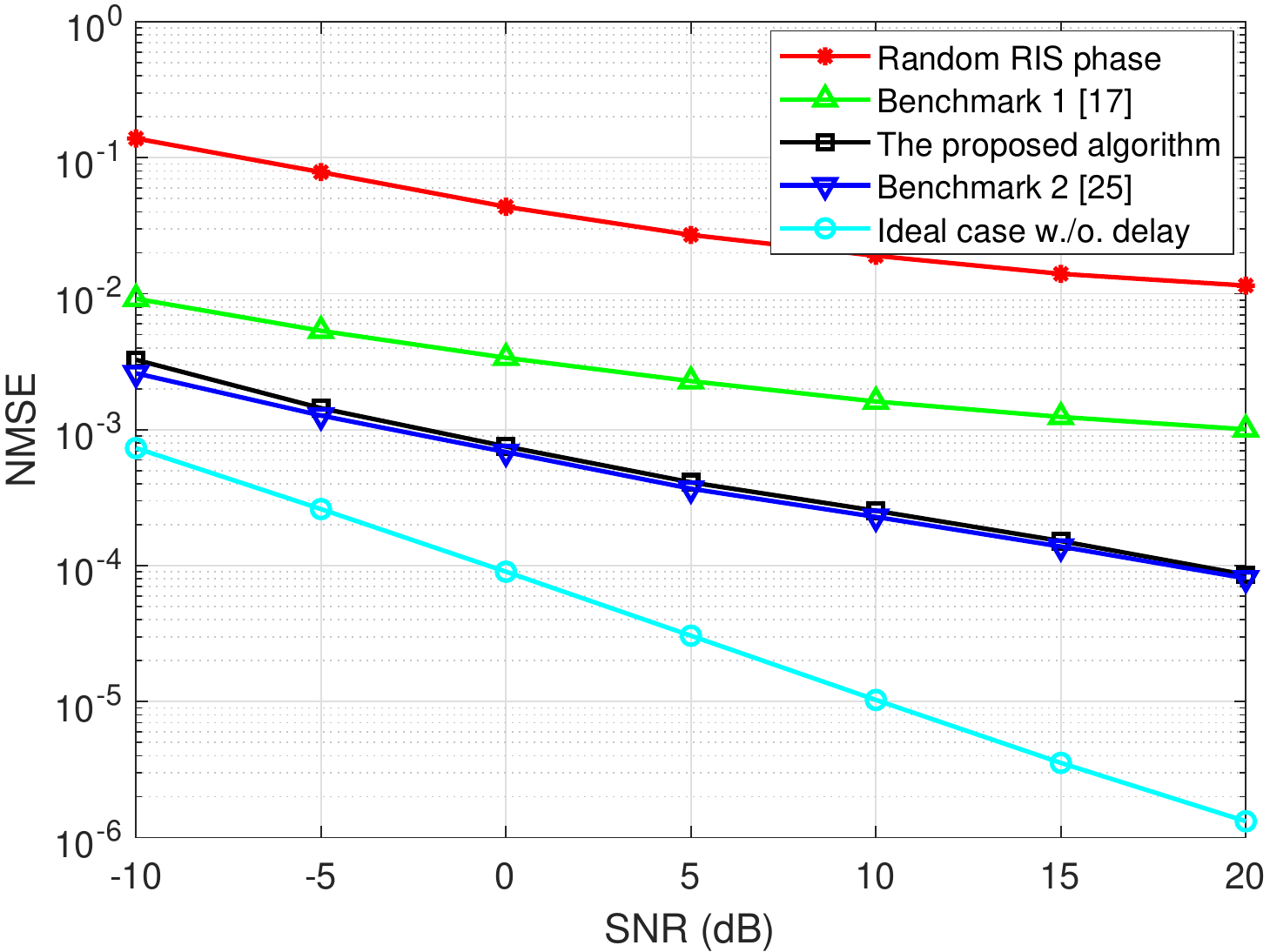}
\caption*{\small{Fig. 5  NMSE performance of different schemes versus SNR over Rayleigh channels ($N=32, K=4$).}}
\label{fig:3}
\end{minipage}
\end{figure}

Next, we test the performance of the cooperative reflection and synchronization design proposed in this paper. For demonstration purposes, we employ the accelerated scheme, i.e., Algorithm 3, to design the RIS reflection matrices (the comparision between Algorithm 2 and Algorithm 3 will be provided later). We mainly compare the proposed algorithm with the following 2 benchmarks:
the conventional scheme where the RISs are designed over
optimistically by assuming perfect timing synchronization (labeled `Benchmark 1'), i.e., the green line, by aligning the phase of the RISs with that of the estimated cascaded channel $\hat{\bf h}_{\text{eq}}$ [17], and the
synchronization design proposed by [25] where perfect timing and channel information is assumed to be known as \emph{a prior} (labeled `Benchmark 2'). On the one hand, it is observed from Fig. 5 that the proposed algorithm outperforms Benchmark 1 by orders-of-magnitude. This verifies that the proposed algorithm can effectively mitigate the ISI and the signal-to-interference-plus-noise ratio loss due to asynchronization in multiple-RIS-assisted systems in terms of NMSE performance. On the other hand, the proposed algorithm provides very close performance to Benchmark 2, i.e., the perfect case without estimation errors, confirming that the proposed estimator is efficient and reliable.
\begin{figure}
\begin{minipage}[t]{0.5\linewidth}
\centering
\includegraphics[width=3.5in]{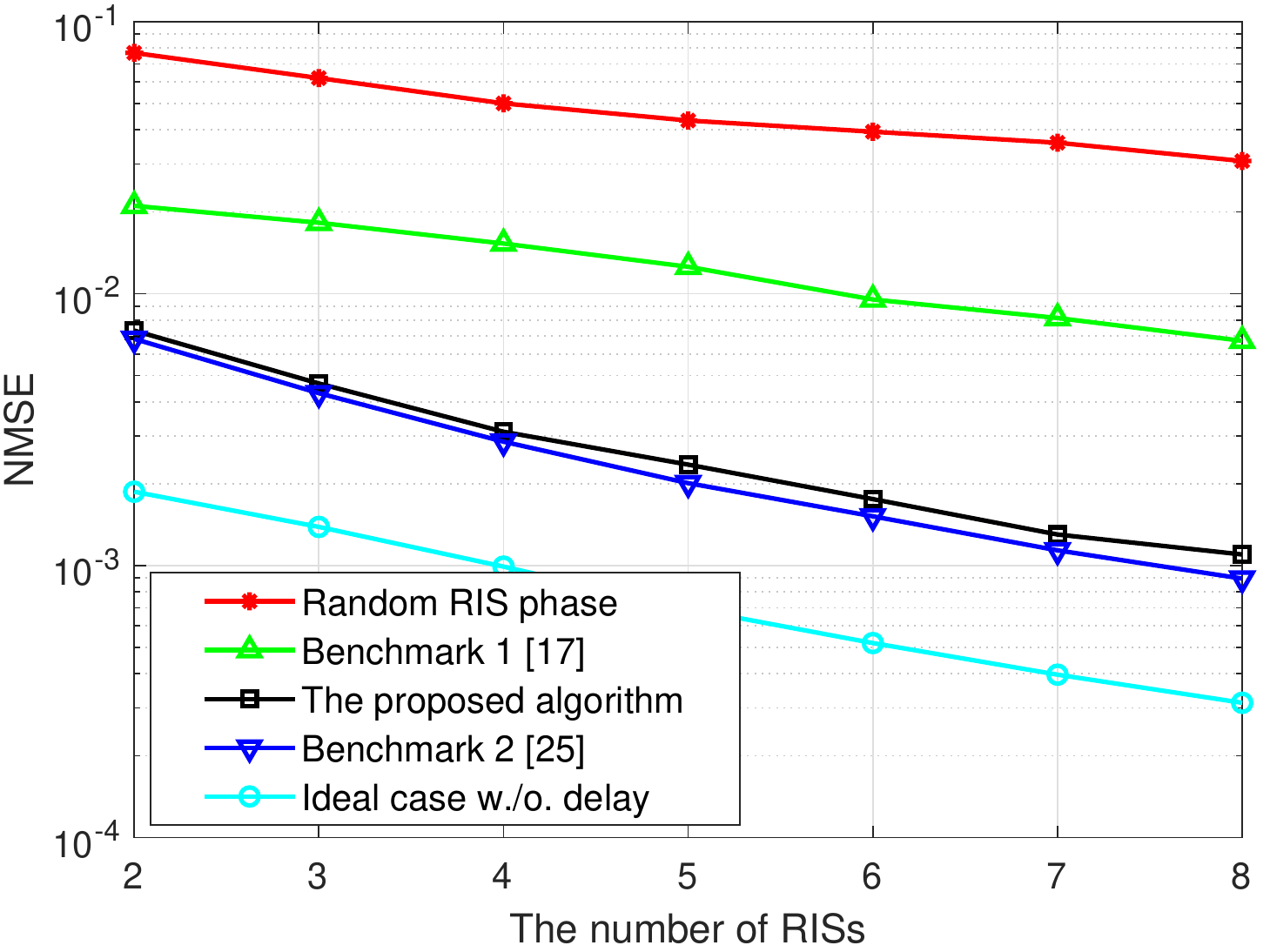}
\caption*{\small{Fig. 6  NMSE performance of different schemes versus the number of RISs over Rayleigh channels ($N=8$).}}
\label{fig:2}
\end{minipage}%
\begin{minipage}[t]{0.5\linewidth}
\centering
\includegraphics[width=3.5in]{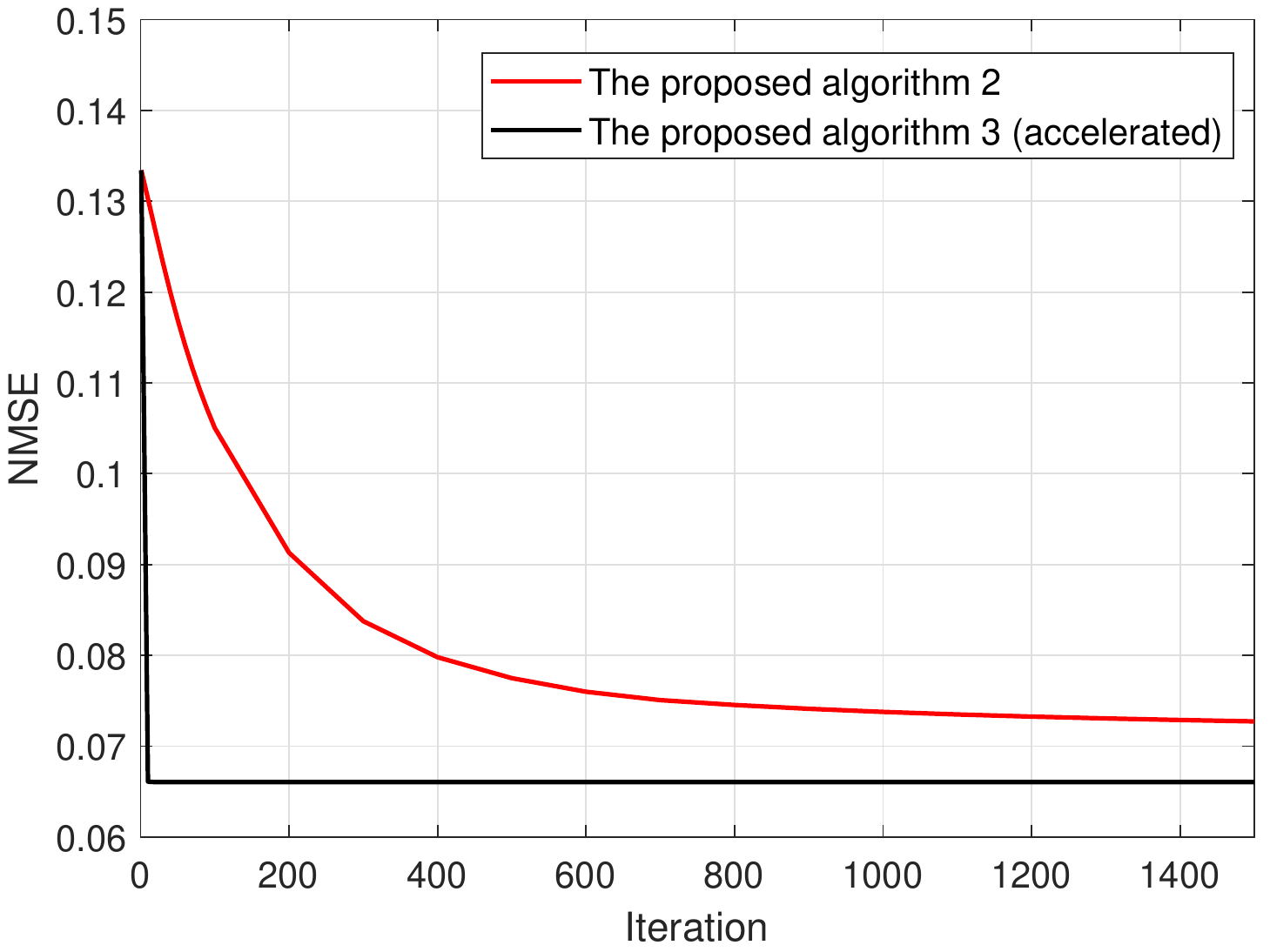}
\caption*{\small{Fig. 7  Convergence of algorithms for the cooperative reflection and synchronization design ($N=4, K=2,$ and SNR = 0 dB).}}
\label{fig:3}
\end{minipage}
\end{figure}

In order to investigate the performance advantage of the proposed design versus the number of RISs, $K$, we plot in Fig. 6 the NMSE of different schemes when SNR is set to 0 dB. According to Fig. 6, the NMSE performance of all schemes monotonically decreases with the number of RISs, while the performance gain achieved by the proposed design gradually increases as $K$ grows compared to random RIS phase design and Benchmark 1 which naively assumes perfect timing synchronization. This indicates that the proposed cooperative reflection and synchronization design is able to better exploit the DoF brought by the multiple RISs for network reliability enhancement. In addition, it is observed that the proposed design incurs limited degradation compared with Benchmark 2 for arbitrary configuration, hence providing relevant guidelines in practical system implementation.

Finally, we experiment with Algorithm 2 and Algorithm 3 to show their convergence
properties. Fig. 7 illustrates the objective values with respect to algorithm iterations. It is shown that both algorithms converge monotonically to a stationary point while Algorithm 3 outperforms Algorithm 2 in terms of both NMSE performance and the convergence rate. This is due to that the continuation property of SQUAREM (Algorithm 3) enables a simple steplength backtracking strategy to obtain global convergence and diminish all the error components. As a result, Algorithm 3 provides significantly more accurate solutions than Algorithm 2.
\subsection{mmWave Channels}
While Rayleigh fading is popularly used in the literatures and has been well regarded as a good model for characterizing typical massive MIMO performance, it is more advisable to consider the narrowband mmWave channel model to depict the nature of high-frequency propagations, e,g., the high free-space pathloss and channel sparsity. Based on the extended Saleh-Valenzuela model, the clustered channel model has been well acknowledged as an accurate model to capture the limited scattering feature of the mmWave channel as discussed in [4]. Specifically, for the channel between ${\mathbb R}_k$ and ${\mathbb D}$, it can be characterized by the sum of all propagation paths that are contributed by limited scatters, which is expressed as
\begin{equation}\label{eq:mmwave}
\mathbf{h}_{k}^{\mathrm{H}}=\sqrt{\frac{N}{N_{p}^k}} \sum_{l=1}^{N_{p}^k} \lambda_{k}^{l}{\bm\alpha}_k^{H}\left(\phi_{k,a}^{l},\phi_{k,e}^{l}\right),\quad\forall k\in\cal K,
\end{equation}
where $N_p^k$ denotes the number of propagation paths between ${\mathbb R}_k$ and ${\mathbb D}$. $\lambda_k^l\sim {\cal {CN}}(0,1)$ is the complex gain of the $l$th path in ${\bf h}_k$ and $\phi _{k,a}^l$ and $\phi_{k,e}^l$ are the azimuth and elevation angle-of-departure corresponding to this path, respectively. ${\bm\alpha}_k(\phi _{k,a},\phi_{k,e})$ is the array response vector of ${\mathbb R}_k$, which is given by
\begin{equation}\label{eq:URA}
\!{\bm\alpha}_k(\phi _{k,a},\phi_{k,e})\!=\!\!\frac{1}{\sqrt{N}}\!\left[\!1,\!\cdots\!,\!e^{\jmath k d(m \sin (\phi _{k,a})\sin (\phi _{k,e})+n \cos (\phi _{k,e}))}\!\cdots\!,\! e^{\jmath k d((N_x-1) \sin (\phi _{k,a}) \sin (\phi _{k,e})+(N_y-1) \cos (\phi _{k,e}))}\right]^{T}\!\!\!\!\!,\!\!\!
\end{equation}
where $0\leq m\leq N_x$ and $0\leq n\leq N_y$ are the azimuth and elevation indices of a reflecting element respectively. For convenience, the number of paths corresponding to different RISs are assumed to be the same, i.e., $N_p^1=N_p^2=\cdots=N_p^K=10$.

For the purpose of coverage extension in mmWave communications that are highly sensitive to blockages, the RISs should be intuitively deployed at locations with clear line-of-sight (LoS) paths from the source. Hence, it is assumed that the ${\mathbb S}$-${\mathbb R}_k$ channel, ${\bf f}_k$, is dominated by the LoS link and we characterize it by a rank-one geometric model expressed as
\begin{figure}
\begin{minipage}[t]{0.5\linewidth}
\centering
\includegraphics[width=3.5in]{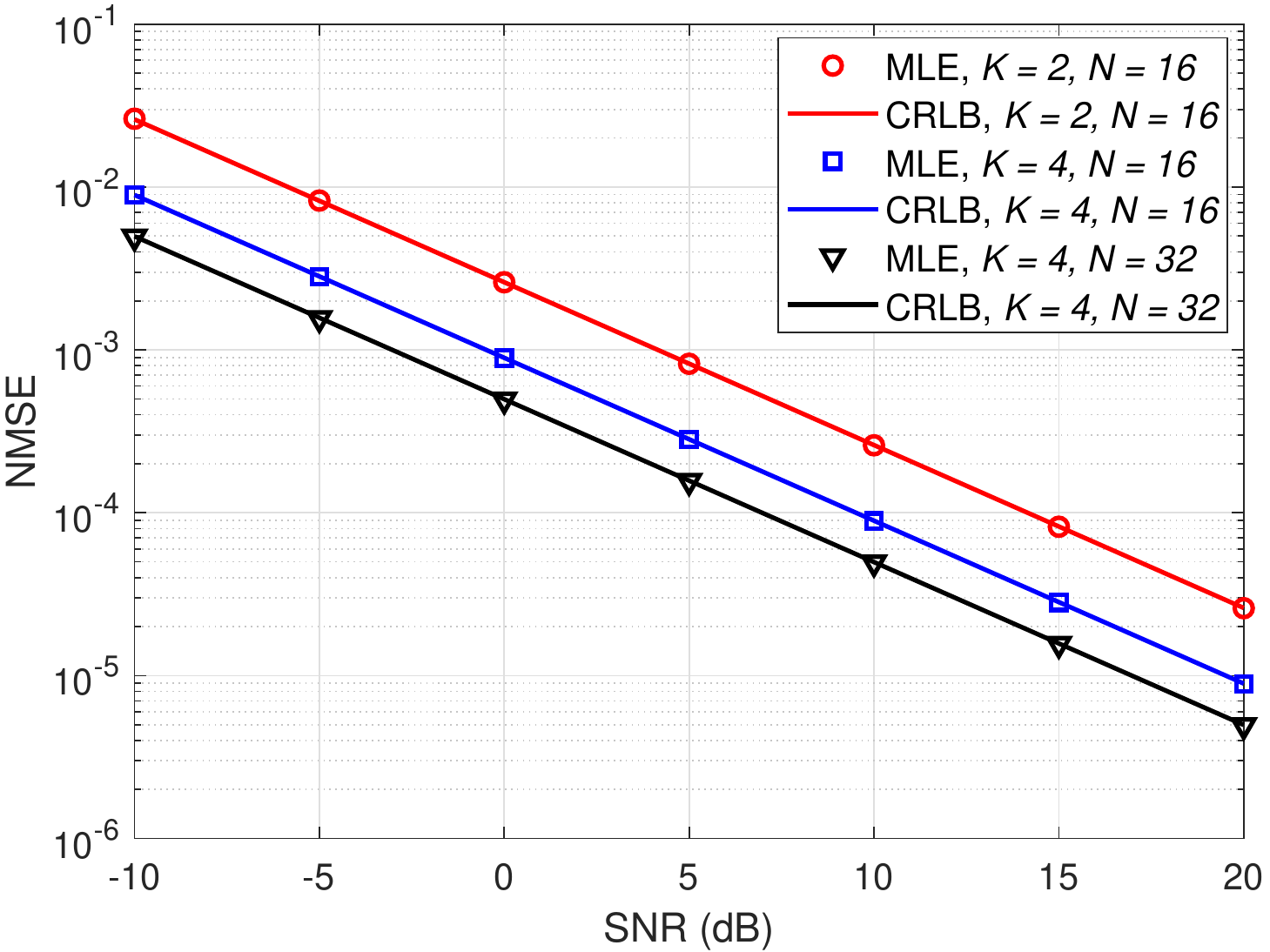}
\caption*{\small{Fig. 8  Estimation performance of the proposed MLE for the cascaded channels ${\bf h}_{\text{eq}}$ over mmWave channels.}}
\label{fig:2}
\end{minipage}%
\begin{minipage}[t]{0.5\linewidth}
\centering
\includegraphics[width=3.5in]{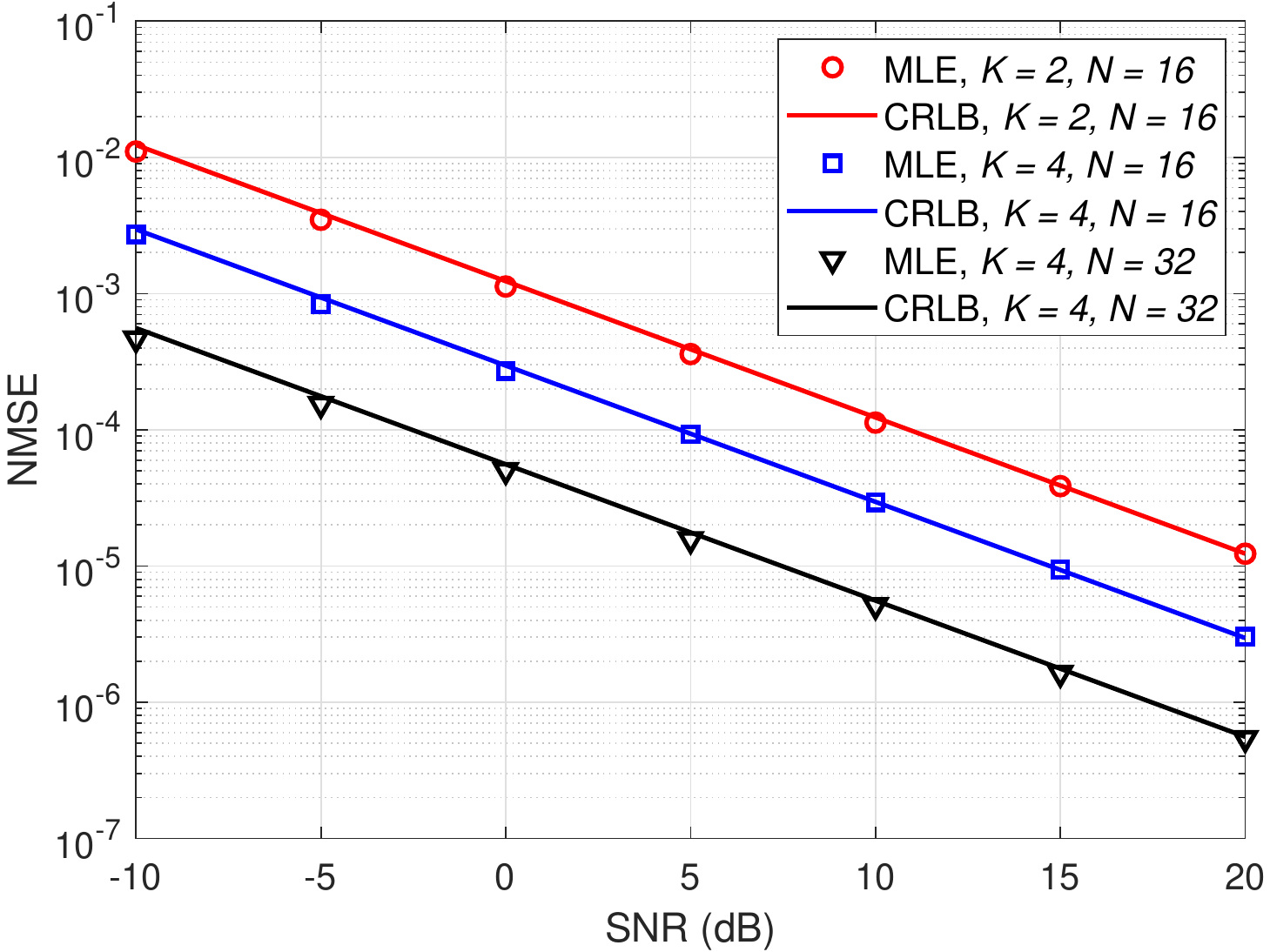}
\caption*{\small{Fig. 9  Estimation performance of the proposed MLE for the timing offsets $\bm\epsilon$ over mmWave channels.}}
\label{fig:3}
\end{minipage}
\end{figure}
\begin{equation}\label{eq:fk}
{\bf f}_k=\sqrt N\lambda_k{\bm\alpha}_k(\vartheta _k^a,\vartheta_k^e),\quad\forall k\in\cal K,
\end{equation}
where $\lambda_k\sim {\cal {CN}}(0,1)$ is the complex gain of the LoS link between ${\mathbb S}$ and ${\mathbb R}_k$ and $\vartheta _k^a$ and $\vartheta_k^e$ respectively correspond to the azimuth and elevation angle-of-arrival of this LoS link.

We first compare the NMSE performance of the proposed estimator and its corresponding CRLB over mmWave channels in Fig. 8 and Fig. 9. We consider the same settings as Fig. 3 and Fig. 4 and similar observations can be also found. Apparently, it is shown in Fig. 8 that the estimation errors of the cascaded channels coincide with the corresponding CRLBs at all SNRs since the values of the channel parameters do not assume any priori finite range. It is obvious in Fig. 9 that on the contrary to the channel estimates, the derived CRLB of the timing offsets can touch the estimation NMSE when SNR is high and deviates from it at low SNRs due to the fact that the given limited range on the timing offsets functions as prior information during estimation.

Then we test the performance of our proposed cooperative reflection and synchronization design over mmWave channels in Fig. 10 and Fig. 11.
\begin{figure}
\begin{minipage}[t]{0.5\linewidth}
\centering
\includegraphics[width=3.5in]{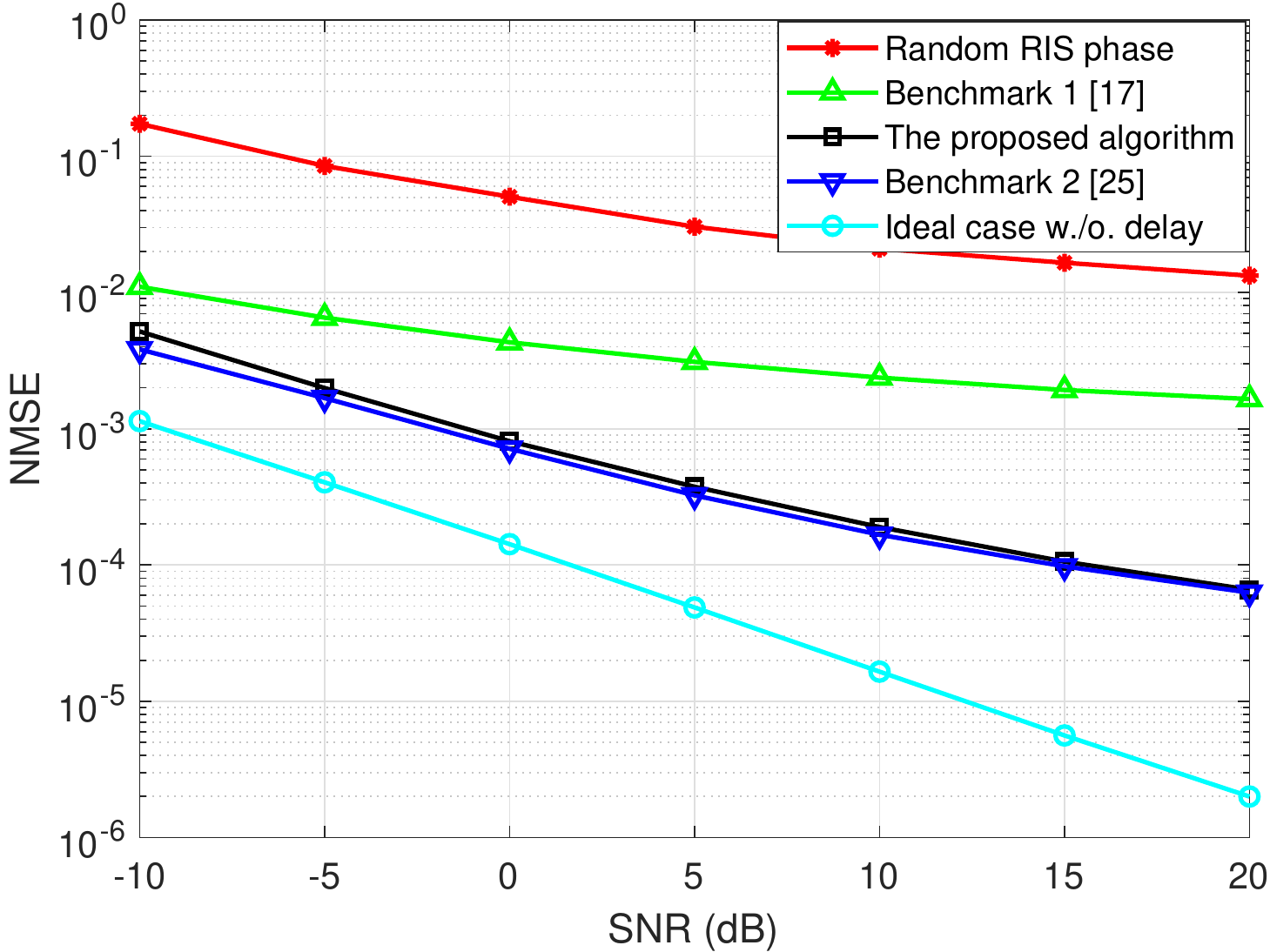}
\caption*{\small{Fig. 10  NMSE performance of different schemes versus SNR over mmWave channels ($N=32, K=4$).}}
\label{fig:2}
\end{minipage}%
\begin{minipage}[t]{0.5\linewidth}
\centering
\includegraphics[width=3.5in]{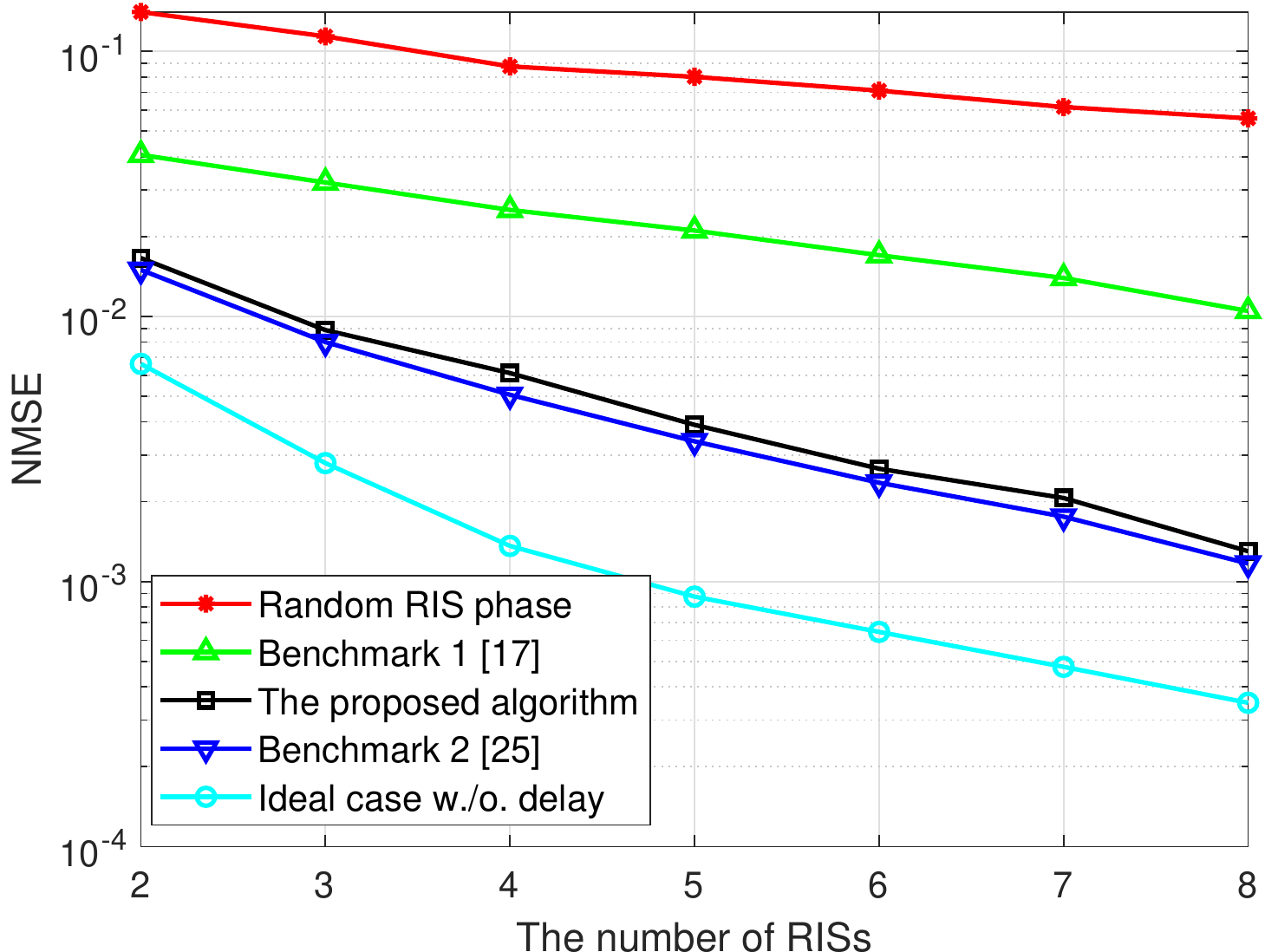}
\caption*{\small{Fig. 11  NMSE performance of different schemes versus the number of RISs over mmWave channels ($N=8$).}}
\label{fig:3}
\end{minipage}
\end{figure}

We adopt the same settings as in Fig. 5 and Fig. 6. Apparently, the proposed algorithm enjoys a satisfactory performance in both Rayleigh and mmWave channels. From Fig. 10, it is shown that the performance of different schemes all increases with SNR while the performance gap between the proposed design and the conventional RIS optimization, i.e., Benchmark 1, also grows with the SNR. Fig. 11 displays the effect of the number of RISs on the NMSE performance using various transmission schemes. As can be seen, the proposed design always exhibits better performance for any $K$ compared with Benchmark 1. In addition, in both Fig. 10 and Fig. 11, the proposed design approaches the synchronized case with perfect timing and channel information, i.e., Benchmark 2.
\section{Conclusion}
In this paper, the problem of joint timing offset and cascaded channel
estimation was studied, and furthermore cooperative reflection and synchronization design for the distributed RIS system has been presented. First, we derived the MLEs and CRLBs for the estimation of timing offsets and channel parameters. Then, aiming at minimizing the MSE of the recovered data at the destination, we jointly optimized the RIS reflection matrices and the timing offset equalizer by taking into account the impact of estimation errors. The formulated problem turns out nonconvex with the unit-modulus constraint imposed on the
RIS reflection coefficients. We develop efficient algorithms to attain a high-quality suboptimal solution to it with convergence guaranteed. Numerical results well demonstrate the accuracy of our theoretical analysis and validate the superiority of the proposed joint estimation and synchronization algorithm.
\begin{appendices}
\section{Useful Lemmas}
\emph{Lemma 1:} Given a matrix ${\bf N}=\left[{\bf n}_1,{\bf n}_2,\cdots,{\bf n}_N\right]\in{\mathbb C}^{M\times N}$ and a vector ${\bf x}=\left[x_1,x_2,\cdots,x_N\right]^T\in{\mathbb C}^{N\times 1}$, where ${\bf n}_l$ is the $l$th column of ${\bf N}$ and $x_l$ is the $l$th element of $\bf x$, $\forall l\in{\cal L}$. If the $l$th element of $\bf x$ is zero, then it gives
\begin{equation}\label{eq:lemma1}
{\bf N}{\bf x} = \hat{\bf N}\hat{\bf x},
\end{equation}
where $\hat{\bf N}=\left[{\bf n}_1,\cdots,{\bf n}_{l-1},{\bf n}_{l+1},\cdots,{\bf n}_N\right]\in{\mathbb C}^{M\times(N-1)}$ and $\hat{\bf x}=\left[x_1,\cdots,x_{l-1},x_{l+1},\cdots,x_N\right]^T\in{\mathbb C}^{(N-1)\times 1}$.
\begin{proof}
This is directly proved by exploiting the fact that zero elements do not contribute to matrix multiplications.
\end{proof}
\emph{Lemma 2:} For an arbitrary square matrix ${\bf X}\in{\mathbb C}^{KL\times KL}$, we have $\mathbb{E}_{{\bm\delta}\left(\bm\epsilon\right)}\left\{{\bf A}(\bm\epsilon){\bf X}{\bf A}(\bm\epsilon)^H\right\}\simeq{\bf A}(\hat{\bm\epsilon}){\bf X}{\bf A}(\hat{\bm\epsilon})^H$ where $\hat{\bm\epsilon}$ is the MLE of $\bm\epsilon$ and $\bm\delta(\bm\epsilon)={\bm\epsilon}-\hat{\bm\epsilon}$ denotes the estimation error.
\begin{proof}
To begin with, we expand ${\bf A}(\epsilon_k)$ with its Taylor series around the estimate $\hat\epsilon_k$ as
\begin{equation}\label{eq:taylor}
 {\bf A}(\epsilon_k)= {\bf A}(\hat\epsilon_k)+{\bf D}({\hat\epsilon_k})\delta(\epsilon_k)+o\left(\delta^2(\epsilon_k)\right),\quad\forall k\in{\cal K},
\end{equation}
where $\delta(\epsilon_k)=\epsilon_k-\hat\epsilon_k$ and the symbol $o\left(\delta^2(\epsilon_k)\right)$ denotes the terms with orders higher than $\delta^2(\epsilon_k)$. Next, denote ${\bf X}[i,j]$ as the submatrix of $\bf X$ with corresponding rows from $(i-1)L+1$ to $iL$ and columns from $(j-1)L+1$ to $jL$, $\forall i\in{\cal K}$ and $\forall j\in{\cal K}$. Then based on the asymptotic distribution of $\delta(\epsilon_k)$ in (\ref{eq:distribution}) and the above Taylor expansion, we get
\begin{align}\label{eq:lemma2}
\mathbb{E}_{{\bm\delta}\left(\bm\epsilon\right)}\left\{{\bf A}(\bm\epsilon){\bf X}{\bf A}(\bm\epsilon)^H\right\}&\overset{(\text a)}=\mathbb{E}_{{\bm\delta}\left(\bm\epsilon\right)}\!\left\{\!\sum\limits_{i,j=1}^{K}{\delta({\epsilon_j}){\bf A}(\hat\epsilon_i){\bf X}[{i,j}]{\bf D}(\hat\epsilon_j)^H}\!\right\}+\mathbb{E}_{{\bm\delta}\left(\bm\epsilon\right)}\!\left\{\!\sum\limits_{i,j=1}^{K}{\delta(\epsilon_i){\bf D}(\hat\epsilon_i){\bf X}[{i,j}]{\bf A}(\hat\epsilon_j)^H}\!\right\}\nonumber\\
&\quad+\sum\limits_{i,j=1}^{K} {{\bf A}(\hat\epsilon_i){\bf X}[i,j]{\bf A}(\hat\epsilon_j)^H}+o\left(\delta^2(\epsilon_k)\right)\nonumber\\
&\overset{(\text b)}\simeq {\bf A}(\hat{\bm\epsilon}){\bf X}{\bf A}(\hat{\bm\epsilon})^H,
\end{align}
where (a) is obtained by elaborating (\ref{eq:taylor}) and (b) can be acquired by utilizing the fact that the first two terms in the equality are zero and neglecting the high order terms.
\end{proof}
\emph{Lemma 4 [37]:} The matrix function $f({\bf X}, {\bf Z}) = {\text {tr}}\left({\bf X}^H{\bf Z}^{-1}{\bf X}\right)$ is jointly convex in ${\bf Z}\succ 0$ and $\bf X$.

\emph{Lemma 5 [25]:} Given ${\bf M}\in\mathbb{C}^{N\times N}\succeq{\bf 0},{\bf Z}\in\mathbb{C}^{M\times M}\succeq{\bf 0}$, and any ${\bf X}_{(t)}\in\mathbb{C}^{M\times N}$, the matrix function ${\text {tr}}\left({\bf Z}{\bf X}{\bf M}{\bf X}^H\right)$ is majorized by $-2 \text{Re}\left\{\text{tr}\left(\left(\lambda \bf{X}_{(t)}-\bf{Z} \bf{X}_{(t)} \bf{M}\right)^{H} \bf{X}\right)\right\}+\lambda\|{\bf X}\|_{F}^{2}+$constant for $\lambda=\|{\bf M}\|_1\|{\bf Z}\|_1$.
\section{Proof of \emph{Theorem 1}}
In order to facilitate the derivations, we first partition the FIM $\bf J$ in (\ref{eq:fisher}) as
\begin{equation}\label{eq:partition}
{\bf{J}} = \frac{2}{\sigma^2_t}\left[ {\begin{array}{*{20}{c}}
{{{\bf{J}}_t}}&{{{\bf{U}}^T}}\\
{\bf{U}}&{{{\bf{J}}_h}}
\end{array}} \right],
\end{equation}
with
\begin{align}\label{eq:parts}
{\bf J}_t&={{\text{Re}}\left\{{\bf H}_{\text e}^*{\bf N}_{\text d}(\bm\epsilon)^H{\bf N}_{\text d}(\bm\epsilon){\bf H}_{\text e}^T\right\} }\nonumber\\
{\bf U}&= \left[ {\begin{array}{*{20}{c}}
{{\text{Re}}\left\{{\bf N}(\bm\epsilon)^H{\bf N}_{\text d}(\bm\epsilon){\bf H}_{\text e}^T\right\}}\\
{{\text{Im}}\left\{{\bf N}(\bm\epsilon)^H{\bf N}_{\text d}(\bm\epsilon){\bf H}_{\text e}^T\right\}}
\end{array}} \right]\nonumber\\
{\bf J}_h&=\left[ {\begin{array}{*{20}{c}}
{{\text{Re}}\left\{{\bf N}(\bm\epsilon)^H{\bf N}(\bm\epsilon)\right\}}&{{-\text{Im}}\left\{{\bf N}(\bm\epsilon)^H{\bf N}(\bm\epsilon)\right\}}\\
{{\text{Im}}\left\{{\bf N}(\bm\epsilon)^H{\bf N}(\bm\epsilon)\right\}}&{{\text{Re}}\left\{{\bf N}(\bm\epsilon)^H{\bf N}(\bm\epsilon)\right\}}
\end{array}} \right].
\end{align}
According to the lemma for the inverse of partitioned matrices, we obtain the expression for ${\bf J}^{-1}$ as follows
\begin{equation}\label{eq:inverse}
{\bf J}^{-1}=\frac{\sigma^2_t}{2}\left[\begin{array}{cc}
\left({\bf J}_t-{\bf U}^T{\bf J}_h^{-1}{\bf U}\right)^{-1}&-\left({\bf J}_t-{\bf U}^T{\bf J}_h^{-1}{\bf U}\right)^{-1}{\bf U}^T{\bf J}_h^{-1} \\
-{\bf J}_h^{-1} {\bf U}\left({\bf J}_t-{\bf U}^T{\bf J}_h^{-1}{\bf U}\right)^{-1}&{\bf J}_h^{-1}+{\bf J}_h^{-1}{\bf U}\left({\bf J}_t-{\bf U}^T{\bf J}_h^{-1}{\bf U}\right)^{-1}{\bf U}^T{\bf J}_h^{-1}
\end{array}\right].
\end{equation}
First, note that
\begin{equation}\label{eq:invFh}
{\bf J}_h^{-1}\!\!=\left[\!\!\!{\begin{array}{*{20}{c}}
{{\text{Re}}\left\{\left({\bf N}(\bm\epsilon)^H{\bf N}(\bm\epsilon)\right)^{-1}\right\}}\!&\!{{-\text{Im}}\left\{\left({\bf N}(\bm\epsilon)^H{\bf N}(\bm\epsilon)\right)^{-1}\right\}}\\
{{\text{Im}}\left\{\left({\bf N}(\bm\epsilon)^H{\bf N}(\bm\epsilon)\right)^{-1}\right\}}\!&\!{{\text{Re}}\left\{\left({\bf N}(\bm\epsilon)^H{\bf N}(\bm\epsilon)\right)^{-1}\right\}}
\end{array}}\!\!\right]
=\left[{\begin{array}{*{20}{c}}
{{\text{Re}}\left\{{\bf Z}\right\}}&{{-\text{Im}}\left\{{\bf Z}\right\}}\\
{{\text{Im}}\left\{{\bf Z}\right\}}&{{\text{Re}}\left\{{\bf Z}\right\}}
\end{array}} \right],
\end{equation}
with ${\bf Z}\triangleq\left({\bf N}(\bm\epsilon)^H{\bf N}(\bm\epsilon)\right)^{-1}$. Then, it is easy to prove the following equation by elaborating the expression of ${\bf U}$ and ${\bf J}_h^{-1}$ in (\ref{eq:parts}) and (\ref{eq:invFh}). It follows
\begin{equation}\label{eq:UFU}
\!{\bf U}^T{\bf J}_h^{-1}{\bf U}\!=\!{\text{Re}}\!\left\{{\bf H}_{\text e}^*{\bf N}_{\text d}(\bm\epsilon)^H{\bf N}(\bm\epsilon){\bf Z}{\bf N}(\bm\epsilon)^H{\bf N}_{\text d}(\bm\epsilon){\bf H}_{\text e}^T\right\},\!
\end{equation}
which further implies that
\begin{equation}\label{eq:upsilon}
\left({\bf J}_t\!-\!{\bf U}^T\!{\bf J}_h^{-1}{\bf U}\right)^{-1}\!\!\!\!\!=\!\!\left({\text{Re}}\left\{\!{\bf H}_{\text e}^*{\bf N}_{\text d}(\bm\epsilon)^H{\bf P}(\bm\epsilon){\bf N}_{\text d}(\bm\epsilon){\bf H}_{\text e}^T\right\}\!\right)^{-1}\!\!\triangleq\!\!\bm\Upsilon.
\end{equation}
Moreover, it holds that
\begin{equation}\label{eq:FhU}
{\bf J}_h^{-1}{\bf U}\!=\!\left[\!\!\!{\begin{array}{*{20}{c}}
{{\text{Re}}\left\{{\bf Z}\right\}}\!\!\!&\!{{-\text{Im}}\left\{{\bf Z}\right\}}\\
{{\text{Im}}\left\{{\bf Z}\right\}}\!\!\!&\!{{\text{Re}}\left\{{\bf Z}\right\}}
\end{array}}\!\!\!\right]\left[\!\!\!{\begin{array}{*{20}{c}}
{{\text{Re}}\left\{{\bf N}(\bm\epsilon)^H{\bf N}_{\text d}(\bm\epsilon){\bf H}_{\text e}^T\right\}}\\
{{\text{Im}}\left\{{\bf N}(\bm\epsilon)^H{\bf N}_{\text d}(\bm\epsilon){\bf H}_{\text e}^T\right\}}
\end{array}}\!\!\!\right]=\left[\!\!\!{\begin{array}{*{20}{c}}
{{\text{Re}}\left\{{\bf Z}{\bf N}(\bm\epsilon)^H{\bf N}_{\text d}(\bm\epsilon){\bf H}_{\text e}^T\right\}}\\
{{\text{Im}}\left\{{\bf Z}{\bf N}(\bm\epsilon)^H{\bf N}_{\text d}(\bm\epsilon){\bf H}_{\text e}^T\right\}}
\end{array}}\!\!\!\right]=\left[\!\!\!{\begin{array}{*{20}{c}}
{{\text{Re}}\{{\bf V}\}}\\
{{\text{Im}}\{{\bf V}\}}
\end{array}}\!\!\!\right],
\end{equation}
where ${\bf V}\triangleq{\bf Z}{\bf N}(\bm\epsilon)^H{\bf N}_{\text d}(\bm\epsilon){\bf H}_{\text e}^T$.

By substituting (\ref{eq:upsilon}) and (\ref{eq:FhU}) into (\ref{eq:inverse}), we have the expression of ${\bf J}^{-1}$ as follows
\begin{equation}\label{eq:inverse1}
{\bf J}^{-1}=\frac{\sigma^2_t}{2}\left[ {\begin{array}{*{20}{c}}
{\bm\Upsilon}&{-\bm\Upsilon\text{Re}\{{\bf V}^H\}}&{\bm\Upsilon\text{Im}\{{\bf V}^H\}}\\
-{\text{Re}\{{\bf V}\}\bm\Upsilon}&{\text{Re}\{{\bf V}\}\bm\Upsilon\text{Re}\{{\bf V}^H\}+\text{Re}\{{\bf Z}\}}&-{\text{Re}\{{\bf V}\}\bm\Upsilon\text{Im}\{{\bf V}^H\}-\text{Im}\{{\bf Z}\}}\\
-\text{Im}\{\bf V\}\bm\Upsilon&{\text{Im}\{{\bf V}\}\bm\Upsilon\text{Re}\{{\bf V}^H\}+\text{Im}\{{\bf Z}\}}&-{\text{Im}\{{\bf V}\}\bm\Upsilon\text{Im}\{{\bf V}^H\}+\text{Re}\{{\bf Z}\}}
\end{array}} \right].
\end{equation}
Then, let ${\bm{\tilde\xi}}\triangleq\left[{\bm\epsilon}^T,{\bf h}_{\text{eq}}^T\right]^T$ and note that
\begin{equation}\label{eq:rtoi}
{\bm{\tilde\xi}}=\left[ \begin{array}{l}
\bm\epsilon\\
{\bf h}_{\text{eq}}
\end{array} \right] = \underbrace{\left[ {\begin{array}{*{20}{c}}
{\bf I}_K&{\bf 0}&{\bf 0}\\
{\bf 0}&{\bf I}_{NK}&\jmath{\bf I}_{NK}
\end{array}} \right]}_{\triangleq{\bm\Gamma}}{\bm\xi}.
\end{equation}
The CRLB matrix of the complex-valued parameters is obtained by inverting the FIM $\bf J$ as
\begin{equation}\label{CRLB}
{\bf C}\left(\bm\epsilon,{\bf h}_{\text{eq}}\right)\!=\!{\bm\Gamma}{\bf J}^{-1}{\bm\Gamma}^H\!\!=\!\!\frac{\sigma^2_t}{2}\left[\!\!{\begin{array}{*{20}{c}}
{\bm\Upsilon}&-{\bm\Upsilon}{\bf V}^H\\
-{\bf V}{\bm\Upsilon}&2{\bf Z}+{\bf V}{\bm\Upsilon}{\bf V}^H
\end{array}}\!\!\right].\!
\end{equation}
Therefore, the CRLBs for $\bm\epsilon$ and ${\bf h}_{\text{eq}}$ are respectively given by
\begin{align}\label{eq:crlbs}
{\bf C}(\bm\epsilon)&=\frac{\sigma _t^2}{2}{\bm\Upsilon}=\frac{\sigma _t^2}{2}\left({\text{Re}}\left\{{\bf H}_{\text e}^*{\bf N}_{\text d}(\bm\epsilon)^H{\bf P}(\bm\epsilon){\bf N}_{\text d}(\bm\epsilon){\bf H}_{\text e}^T\right\}\right)^{-1}\nonumber\\
{\bf C}\left({\bf h}_{\text{eq}}\right)&=\frac{\sigma _t^2}{2}\left(2{\bf Z}+{\bf V}{\bm\Upsilon}{\bf V}^H\right),\nonumber\\
&=\frac{\sigma _t^2}{2}\!\left(\!2\left({\bf N}(\bm\epsilon)^H{\bf N}(\bm\epsilon)\right)^{-1}\!\!\!+\!\left({\bf N}(\bm\epsilon)^H{\bf N}(\bm\epsilon)\right)^{-1}\!{\bf N}(\bm\epsilon)^H{\bf N}_{\text d}(\bm\epsilon){\bf H}_{\text e}^T\left({\text{Re}}\left\{{\bf H}_{\text e}^*{\bf N}_{\text d}(\bm\epsilon)^H{\bf P}(\bm\epsilon){\bf N}_{\text d}(\bm\epsilon){\bf H}_{\text e}^T\right\}\right)^{-1}\right.\nonumber\\
&\quad\left.\times{\bf H}_{\text e}^*{\bf N}_{\text d}(\bm\epsilon)^H{\bf N}(\bm\epsilon)\left({\bf N}(\bm\epsilon)^H{\bf N}(\bm\epsilon)\right)^{-1}\right),
\end{align}
which proves the theorem.
\section{Calculations of MSE$({\bm\Psi},{\bf G})$}
By substituting (\ref{eq:yd}) into (\ref{eq:problema}), we have
\begin{align}\label{eq:MSE0}
\text{MSE}({\bm \Psi},{\bf G})&=\mathbb{E}_{{\bm\delta}\left(\bm\epsilon\right),{\bm\delta}\left({\bf h}_{\text{eq}}\right),{\bf s}_d,{\bf v}}\left\{ {\left\|{\bf G}\left({\bf A}({\bm\epsilon}){\bm \Psi}{\bf H}_{\text{eq}}{\bf s}_d+{\bf v}\right) - {\bf T}({\bm{\eta }}){\bf s}_d \right\|^2 }\right\}\nonumber\\
&={\text{tr}}\left\{{\bf G}\mathbb{E}_{{\bm\delta}\left(\bm\epsilon\right)}\left\{{\bf A}(\bm\epsilon){\bm \Psi}\mathbb{E}_{{\bm\delta}\left({\bf h}_{\text{eq}}\right)}\left\{{\bf H}_{\text{eq}}{\bf H}_{\text{eq}}^H\right\}{\bm\Psi}^H{\bf A}(\bm\epsilon)^H\right\}{\bf G}^H\right\}+{\text{tr}}\left\{{\bf G}{\bf R}_v{\bf G}^H\right\}\nonumber\\
&\quad-2{\text{Re}}\left\{{\text{tr}}\left\{{\bf G}\mathbb{E}_{{\bm\delta}\left(\bm\epsilon\right)}\left\{{\bf A}(\bm\epsilon)\right\}{\bm \Psi}\mathbb{E}_{{\bm\delta}\left({\bf h}_{\text{eq}}\right)}\left\{{\bf H}_{\text{eq}}\right\}{\bf T}^H({\bm{\eta }})\right\}\right\}+{\text{tr}}\left\{{\bf T}({\bm{\eta }}){\bf T}^H({\bm{\eta }})\right\},
\end{align}
where ${\bf R}_v\triangleq\mathbb{E}\left\{{\bf v}{\bf v}^H\right\}$.

Note that the expectation operations in (\ref{eq:MSE0}) is difficult to be handled. For this reason, we manipulate it by utilizing the statistical information of the estimation errors of timing offsets and channel parameters in the following.
\begin{itemize}
  \item \textbf{Channel Parameters:} Using the fact that ${\bf h}_{\text{eq}} = {\hat{\bf h}_{\text{eq}}}+{\bm\delta}\left({\bf h}_{\text{eq}}\right)$, we can write
\begin{equation}\label{eq:Heq}
{\bf H}_{\text{eq}} = {\hat{\bf H}_{\text{eq}}}+{\bm\delta}\left({\bf H}_{\text{eq}}\right),
\end{equation}
where $\hat{\bf H}_{\text{eq}}=\hat{\bf h}_{\text{eq}}\otimes{\bf I}_L$, ${\bm\delta}\left({\bf H}_{\text{eq}}\right)={\bm\delta}\left({\bf h}_{\text{eq}}\right)\otimes{\bf I}_L$. Based on the asymptotic Gaussian distribution in (\ref{eq:distribution}) that ${\bm\delta}\left({\bf h}_{\text{eq}}\right)\sim{\cal {CN}}\left({\bf 0},{\bf C}\left({\bf h}_{\text{eq}}\right)\right)$, the expectations in (\ref{eq:MSE0}) over ${\bm\delta}\left({\bf h}_{\text{eq}}\right)$ can be evaluated as
\begin{align}\label{eq:EHeq}
\mathbb{E}_{{\bm\delta}\left({\bf h}_{\text{eq}}\right)}\left\{{\bf H}_{\text{eq}}\right\}&={\hat{\bf H}_{\text{eq}}},\nonumber\\
\mathbb{E}_{{\bm\delta}\left({\bf h}_{\text{eq}}\right)}\left\{{\bf H}_{\text{eq}}{\bf H}_{\text{eq}}^H\right\}&={\bf R}_{{\bm\delta}({{\bf H}_{\text{eq}}})}\approx{\bf R}_{{\bm\delta}({\hat{\bf H}_{\text{eq}}})},
\end{align}
where ${\bf R}_{{\bm\delta}({{\bf H}_{\text{eq}}})}={\bf R}_{{\bm\delta}({{\bf h}_{\text{eq}}})}\otimes{\bf I}_L$ with ${\bf R}_{{\bm\delta}({{\bf h}_{\text{eq}}})}={\hat{\bf h}_{\text{eq}}}{\hat{\bf h}^H_{\text{eq}}}+{\bf C}\left({\bf h}_{\text{eq}}\right)$ and ${\bf R}_{{\bm\delta}({\hat{\bf H}_{\text{eq}}})}$ is obtained by substituting ${\hat{\bf H}_{\text{eq}}}$ into the expression of ${\bf R}_{{\bm\delta}({{\bf H}_{\text{eq}}})}$.
  \item \textbf{Timing Parameters:} First, it is easy to obtain $\mathbb{E}_{{\bm\delta}\left(\bm\epsilon\right)}\left\{{\bf A}(\bm\epsilon)\right\}={\bf A}(\hat{\bm\epsilon})$ according to (\ref{eq:distribution}). As for the quadratic term with respect to ${\bf A}(\bm\epsilon)$ in (\ref{eq:MSE0}), by exploiting \emph{Lemma 2} in Appendix A and (\ref{eq:EHeq}), we obtain
\begin{equation}\label{eq:EAe}
\mathbb{E}_{{\bm\delta}\left(\bm\epsilon\right)}\left\{{\bf A}(\bm\epsilon){\bm \Psi}\mathbb{E}_{{\bm\delta}\left({\bf h}_{\text{eq}}\right)}\left\{{\bf H}_{\text{eq}}{\bf H}_{\text{eq}}^H\right\}{\bm\Psi}^H{\bf A}(\bm\epsilon)^H\right\}={\bf A}(\hat{\bm\epsilon}){\bm \Psi}{\bf R}_{{\bm\delta}({\hat{\bf H}_{\text{eq}}})}{\bm\Psi}^H{\bf A}(\hat{\bm\epsilon})^H.
\end{equation}
\end{itemize}
Building upon the above analysis, the MSE expression in (\ref{eq:MSE0}) is finally evaluated as
\begin{align}\label{eq:MSE1}
\text{MSE}({\bm \Psi},{\bf G})&={\text{tr}}\left\{{\bf G}{\bf A}(\hat{\bm\epsilon}){\bm \Psi}{\bf R}_{{\bm\delta}({\hat{\bf H}_{\text{eq}}})}{\bm\Psi}^H{\bf A}(\hat{\bm\epsilon})^H{\bf G}^H\right\}+{\text{tr}}\left\{{\bf G}{\bf R}_v{\bf G}^H\right\}\nonumber\\
&\quad-2{\text{Re}}\left\{{\text{tr}}\!\left\{{\bf G}{\bf A}(\hat{\bm\epsilon}){\bm \Psi}\hat{\bf H}_{\text{eq}}{\bf T}^H\!({\bm{\eta }})\right\}\right\}+{\text{tr}}\left\{{\bf T}({\bm{\eta }}){\bf T}^H({\bm{\eta }})\right\}.
\end{align}
\section{Proof of Lemma 3}
Since ${\bf R}_{{\bm\delta}({\hat{\bf H}_{\text{eq}}})}$ is a positive-definite matrix, we can transform the original MSE expression (\ref{eq:MSE11}) into the following form as
\begin{align}\label{eq:MSE2}
\text{MSE}({\bm \Psi},{\bf G})&={\text{tr}}\left\{{\bf G}{\bf A}(\hat{\bm\epsilon}){\bm \Psi}{\bf R}_{{\bm\delta}({\hat{\bf H}_{\text{eq}}})}^{1/2}{\bf R}_{{\bm\delta}({\hat{\bf H}_{\text{eq}}})}^{H/2}{\bm\Psi}^H{\bf A}(\hat{\bm\epsilon})^H{\bf G}^H\right\}+{\text{tr}}\left\{{\bf G}{\bf R}_v{\bf G}^H\right\}\nonumber\\
&\quad-2{\text{Re}}\left\{{\text{tr}}\left\{{\bf G}{\bf A}(\hat{\bm\epsilon}){\bm \Psi}\hat{\bf H}_{\text{eq}}{\bf T}^H({\bm{\eta }})\right\}\right\}+{\text{tr}}\left\{{\bf T}({\bm{\eta }}){\bf T}^H({\bm{\eta }})\right\}.
\end{align}
Recall the definition of ${\bf A}(\bm\epsilon)$, ${\bm \Psi}\triangleq\text{blk}\left[{\bm\theta}_1^T,{\bm\theta}_2^T,\cdots,{\bm\theta}_K^T\right]\otimes{\bf I}_L$, $\hat{\bf H}_{\text{eq}}=\hat{\bf h}_{\text{eq}}\otimes{\bf I}_L$, and ${\bf R}_{{\bm\delta}({\hat{\bf H}_{\text{eq}}})}^{1/2}={\bf R}_{{\bm\delta}({\hat{\bf h}_{\text{eq}}})}^{1/2}\otimes{\bf I}_L$. Then by using the property of Kronecker product, we have
\begin{align}\label{eq:convert}
{\bf A}(\hat{\bm\epsilon}){\bm \Psi}{\bf R}_{{\bm\delta}({\hat{\bf H}_{\text{eq}}})}^{1/2}&={\bm\Theta}{\bm\Xi},\nonumber\\
{\bf A}(\hat{\bm\epsilon}){\bm \Psi}\hat{\bf H}_{\text{eq}}&={\bm\Theta}{\bm\Pi},
\end{align}
where ${\bm\Theta}\triangleq{\bm\theta}^T\otimes{\bf I}_{L_oQ}$ and
\begin{align*}
\!\begin{split}
\left\{ \begin{array}{l}
\!\!{\bm\Xi}\triangleq\left[{\bf R}_1^T\otimes{\bf A}(\hat\epsilon_1)^T\!,{\bf R}_2^T\otimes{\bf A}(\hat\epsilon_2)^T\!,\cdots,{\bf R}_K^T\otimes{\bf A}(\hat\epsilon_K)^T\right]^T\!\!\\
\!\!{\bm\Pi}\triangleq\left[\hat{\bf h}_{\text{eq},1}^T\!\otimes\!{\bf A}(\hat\epsilon_1)^T\!\!,\hat{\bf h}_{\text{eq},2}^T\!\otimes\!{\bf A}(\hat\epsilon_2)^T\!\!,\cdots\!,\hat{\bf h}_{\text{eq},K}^T\!\otimes\!{\bf A}(\hat\epsilon_K)^T\right]^T\!\!\!
\end{array} \right.
\end{split}
\end{align*}
with ${\bf R}_{{\bm\delta}({\hat{\bf h}_{\text{eq}}})}^{1/2}=\left[{\bf R}_1^T,{\bf R}_2^T,\cdots,{\bf R}_K^T\right]^T$ and $\hat{\bf h}_{\text{eq},k}\in{\mathbb C}^{N\times 1}$ representing the estimate of the cascaded channel corresponding to ${\mathbb R}_k$. Note that this result can be easily obtained by elaborating the left-hand side of the equations exploiting the basic properties of matrix multiplication.

To this end, we substitute (\ref{eq:convert}) into (\ref{eq:MSE2}), which yields a MSE expression in terms of ${\bm\Theta}$ and $\bf G$
\begin{equation}\label{eq:MSE3}
\text{MSE}({\bm \Theta},{\bf G})\!=\!{\text{tr}}\left\{{\bf G}{\bm\Theta}{\bm\Xi}{\bm\Xi}^H{\bm\Theta}^H{\bf G}^H\right\}+{\text{tr}}\left\{{\bf G}{\bf R}_v{\bf G}^H\right\}-2{\text{Re}}\left\{{\text{tr}}\left\{{\bf G}{\bm\Theta}{\bm\Pi}{\bf T}^H\!({\bm{\eta }})\right\}\right\}\!+\!{\text{tr}}\left\{{\bf T}({\bm{\eta }}){\bf T}^H\!({\bm{\eta }})\right\}\!,\nonumber
\end{equation}
which is the desired result in \emph{Lemma 3}.
\section{Proof of Theorem 3}
First, we denote the constraint set of problem $({\cal P}_3)$ by ${\cal S}$. In order to prove the convergence of Algorithm 2, we need to first verify the following four conditions according to [38]:

1) $\overline{\text{MSE}}({\bm\Theta})\geq g\left({\bm\Theta},{\bm\Theta}_{(t)}\right),\forall{\bm\Theta },{\bm\Theta}_{(t)}\in{\cal S}$;

2) $\overline{\text{MSE}}({\bm\Theta}_{(t)})=g\left({\bm\Theta}_{(t)},{\bm\Theta}_{(t)}\right)$;

3) $\overline{\text{MSE}}'\left({\bm\Theta}_{(t)};{\bf d}\right)=g'({\bm\Theta},{\bm\Theta}_{(t)};{\bf d})|_{{\bm\Theta}={\bm\Theta}_{(t)}},\forall{\bf d}$ with ${\bm\Theta}_{(t)}+{\bf d}\in{\cal S}$;

4) $g\left({\bm\Theta},{\bm\Theta}_{(t)}\right)$ is continuous in ${\bm\Theta}$ and ${\bm\Theta}_{(t)}$.

Under the above conditions, every limit point of the sequence $\{{\bm\Theta}_{(t)}\}$ is a locally optimal point of the considered problem in $({\cal P}_3)$ (see [35] for details).

It can be readily confirmed that conditions 1), 2), and 4) hold for $\overline{\text{MSE}}({\bm\Theta})$
and $g\left({\bm\Theta},{\bm\Theta}_{(t)}\right)$. Concerning condition 3), we calculate $\overline{\text{MSE}}'\left({\bm\Theta};{\bf d}\right)$ and $g'({\bm\Theta},{\bm\Theta}_{(t)};{\bf d})$ as
\begin{align}
\overline{\text{MSE}}'\left({\bm\Theta};{\bf d}\right)&=2\text{Re}\!\left\{\text{tr}\left\{\left(-{\bf F}_{(t)}{\bf F}_{(t)}^H{\bm\Theta}_{(t)}{\bm\Xi}{\bm\Xi}^H+{\bf F}_{(t)}{\bf T}(\bm\eta){\bm\Pi}^H\!\right)\!{\bf d}\right\}\right\},\nonumber\\
g'({\bm\Theta},{\bm\Theta}_{(t)};{\bf d})
&=2\text{Re}\left\{\text{tr}\left\{\left(\lambda_{(t)}\left({\bm\Theta}_{(t)}-{\bm\Theta}\right)-{\bf F}_{(t)}{\bf F}_{(t)}^H{\bm\Theta}_{(t)}{\bm\Xi}{\bm\Xi}^H+{\bf F}_{(t)}{\bf T}(\bm\eta){\bm\Pi}^H\right){\bf d}\right\}\right\}.
\end{align}Clearly, we have $\overline{\text{MSE}}'\left({\bm\Theta}_{(t)};{\bf d}\right)=g'({\bm\Theta},{\bm\Theta}_{(t)};{\bf d})|_{{\bm\Theta}={\bm\Theta}_{(t)}}$ and then all the four conditions hold. Therefore, the sequence of the solutions obtained in each iteration will result in a monotonically increasing objective value $\{\overline{\text{MSE}}({\bm\Theta}_{(t)}), t = 1, 2, \cdots\}$ and finally converge to a local optimum of the problem $({\cal P}_3)$.

On the other hand, the optimality of $\bf G$ is guaranteed by the derived closed-form solution in equation (25). Consequently, Algorithm 2 is guaranteed to converge to a local optimum of the original design problem in $({\cal P}_1)$.
\end{appendices}

\end{document}